\documentstyle[12pt,epsfig,axodraw]{article}

\newcommand{\s}{\smallskip}

\newcommand{\nn}{\noindent}
\newcommand{\non}{\nonumber}
\newcommand{\beq}{\begin{eqnarray}}
\newcommand{\eeq}{\end{eqnarray}}
\newcommand{\mx}{m_{\chi^0_i}}
\newcommand{\mxx}{m_{\chi_i}}
\newcommand{\me}{m_{\tilde{e}_{L,R}}}
\newcommand{\ml}{m_{\tilde{\ell}}}
\newcommand{\mxc}{m_{\chi^-_i}}
\newcommand{\mn}{m_{\tilde{\nu}_{L}}}
\newcommand{\ee}{e^+e^-}
\newcommand{\tb}{\tan \beta}

\newcommand{\mgi}{m_{\chi_i}}
\newcommand{\mgj}{m_{\chi_j}}
\newcommand{\msf}{m_{\tilde{f}}}

\newcommand{\cala}[1]{{\cal A}_{_#1}}
\newcommand{\calb}[1]{{\cal B}_{_#1}}
\newcommand{\calc}[1]{{\cal C}_{_#1}}
\newcommand{\cald}[1]{{\cal D}_{_#1}}
\newcommand{\calf}[1]{{\cal F}_{_#1}}

\newcommand{\pn}[1]{\not{p}_#1}

\newcommand{\ctw}{c_W}
\newcommand{\stw}{s_W}
\newcommand{\ctwn}[1]{c^#1_W}
\newcommand{\stwn}[1]{s^#1_W}

\begin{document}

\vspace*{.1cm} 
\baselineskip=17pt

\begin{flushright}
PM/02--09\\
April 2002\\
\end{flushright}

\vspace*{0.9cm}

\begin{center}

{\large\sc {\bf Associated production of sfermions and gauginos}}

\vspace*{0.4cm}

{\large\sc {\bf at high--energy e$^+$e$^-$ colliders:}}

\vspace*{0.4cm}

{\large\sc {\bf the case of selectrons and electronic sneutrinos}}

\vspace{0.7cm}

Aseshkrishna {\sc Datta}, Abdelhak {\sc Djouadi} and Margarete
{\sc M\"uhlleitner}

\vspace{0.7cm}

Laboratoire de Physique Math\'ematique et Th\'eorique, UMR5825--CNRS,\\
Universit\'e de Montpellier II, F--34095 Montpellier Cedex 5, France. 

\end{center} 

\vspace*{1cm} 

\begin{abstract}
\nn We analyze the associated production at future high--energy $e^+e^-$
colliders, of first generation sleptons with neutralinos and charginos in the
modes $e^+ e^-$ and $e \gamma$, in the framework of the Minimal Supersymmetric
extension of the Standard Model. We show that the production rates, in
particular for associated production of right--handed selectrons and the
lightest neutralino which in general is the first accessible kinematically, can
be much larger than the corresponding ones for second and third generation
scalar leptons and for scalar quarks. With the high--luminosities expected at
these colliders, the detection of first generation sleptons with masses
significantly above the kinematical two--body threshold, $\sqrt{s}= 2m_{
\tilde{\ell}}$, is thus possible in favourable regions of the parameter space.  
\end{abstract}

\newpage 

\subsection*{1. Introduction}

In a preceding paper \cite{first}, the associated production of scalar fermions
with neutralinos and charginos at future high--energy $\ee$ colliders has been
analyzed in the context of the Minimal Supersymmetric Standard Model (MSSM). 
For scalar leptons, only the cases of left-- and right--handed smuons, staus
and their partner sneutrinos have been considered. The associated production
of scalar quarks, including  the lightest third generation states,
$\tilde{t}_1$ and $\tilde{b}_1$, with neutralinos, charginos as well as with
gluinos, have also been studied. It has been shown that some of these
three--body processes can have production cross sections sizeable enough to
allow for the possibility of discovering sfermions with masses slightly above
the kinematical threshold for pair production at ${\cal O}($1 TeV) $\ee$
colliders, i.e.~$\sqrt{s}= 2 m_{\tilde{f}}$, in favorable regions of the
supersymmetric (SUSY) parameter space. \s

The case comprising of associated production of first generation sleptons,
left-- and right--handed selectrons $\tilde{e}_{L,R}$ and electronic sneutrinos
$\tilde{\nu}_e$, with charginos and neutralinos is more complicated compared to
the case of second and third generation sleptons. This is mostly due to the
fact that, because of the possibility of the $t$--channel exchange of charginos
and neutralinos in the two--body production processes, $\ee \to \tilde{e}_{L/R}
\tilde{e}_{L/R}$ and $\ee \to \tilde{\nu}_e \tilde{\nu}_e$, respectively, there
are many more contributing Feynman diagrams for the three--body final states,
$\ee \to \ell_e \tilde{\ell}_e \chi$. In addition, there are extra
contributions with diagrams involving the exchange of $\gamma, Z, W$ gauge
bosons in the $t$--channel. Particularly important are those with $t$--channel
$\gamma$ exchange, which for almost real photons, lead to poles that need to 
be handled carefully. \s

In the present paper, we extend the analysis of Ref.~\cite{first} to include
the  case of associated production of first generation scalar leptons with the
lighter charginos and neutralinos. We first investigate the associated 
production of selectrons and sneutrinos with the lighter neutralinos 
$\chi_{1,2}^0$ or chargino $\chi_{1}^\pm$ states in electron--photon collisions
\cite{egamma1,egamma2}
\beq
\gamma \, e_{L,R}^\pm \to \tilde{e}_{L,R}^\pm \, \chi_{1,2}^0 \ \ ,  \ \ 
\gamma \, e_{L,R}^\pm \to \tilde{\nu}_{e} \chi_1^\pm  
\eeq
for real photons coming from the initial electron/positron beams in the original
\cite{WW} and improved \cite{IWW} Weizs\"acker--Williams approximations, as 
well as in the case of an $e\gamma$ collider with the photon generated by 
Compton--back scattering of laser light \cite{laser}. We then analyze the 
three--body processes for selectron and sneutrino productions
\beq
\ee \, \to \,  e^\pm \, \tilde{e}_{L,R}^\mp \, \chi_{1,2}^0 \  &,&  \ 
\ee \, \to \, \nu \; \tilde{e}_{L,R}^\pm  \; \chi_1^\mp \nonumber \\
\ee \, \to \,  \nu_e \  \tilde{\nu}_{e} \ \chi_{1,2}^0 \ &,& \ 
\ee \, \to \, e^\pm \tilde{\nu} \, \chi_1^\mp
\eeq
in the $\ee$ mode of a future linear collider with c.m.~energies
$\sqrt{s}=500$ GeV and 1 TeV.

\subsection*{2. Production in the e$\gamma$ mode} 

The associated production of selectrons with neutralinos and sneutrinos with
charginos in $e^- \gamma$ collisions [the cross sections are the same in $e^+
\gamma$ processes because of CP invariance] occur through the $s$--channel
exchange of electrons and the $t$--channel exchange of  selectrons or
charginos; Fig.~1.  

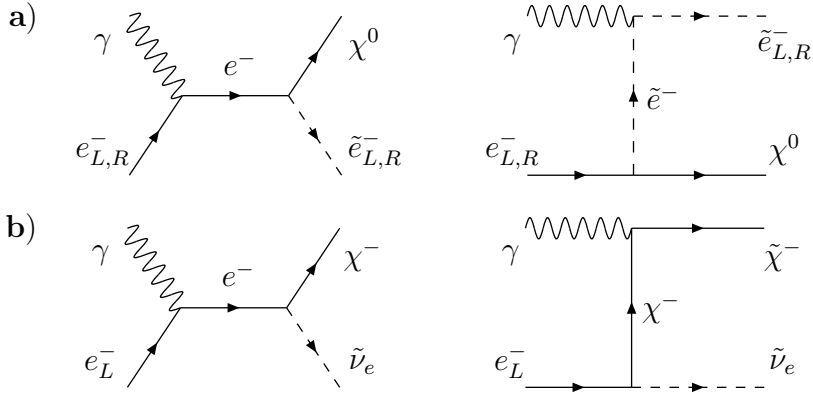
\begin{figure}[htbp]
\vspace*{.3cm}
\hspace*{3cm}
\begin{picture}(1000,180)(0,0)
\Text(-20,180)[]{${\bf a)}$}
\Text(10,130)[]{$e^-_{L,R}$}
\Text(10,170)[]{$\gamma$}
\ArrowLine(20,120)(40,150)
\Photon(40,150)(20,180){4}{6}
\ArrowLine(40,150)(80,150)
\Text(62,163)[]{$e^-$}
\DashArrowLine(80,150)(100,120){4}
\ArrowLine(80,150)(100,180)
\Text(113,130)[]{$\tilde{e}^-_{L,R}$}
\Text(109,170)[]{$\chi^0$}
\Text(165,130)[]{$e^-_{L,R}$}
\Text(165,170)[]{$\gamma$}
\Photon(170,180)(210,180){4}{6}
\ArrowLine(170,120)(210,120)
\DashArrowLine(210,180)(260,180){4}
\Text(222,150)[]{$\tilde{e}^-$}
\ArrowLine(210,120)(260,120)
\DashArrowLine(210,120)(210,180){4}
\Text(268,130)[]{$\chi^0$}
\Text(268,170)[]{$\tilde{e}^-_{L,R}$}
\hspace*{-3mm}
\Text(-20,100)[]{${\bf b)}$}
\Text(10,50)[]{$e^-_{L}$}
\Text(10,90)[]{$\gamma$}
\ArrowLine(20,40)(40,70)
\Photon(40,70)(20,100){4}{6}
\ArrowLine(40,70)(80,70)
\Text(62,83)[]{$e^-$}
\DashArrowLine(80,70)(100,40){4}
\ArrowLine(80,70)(100,100)
\Text(109,50)[]{$\tilde{\nu}_e$}
\Text(109,90)[]{$\chi^-$}
\Text(165,50)[]{$e^-_L$}
\Text(165,90)[]{$\gamma$}
\Photon(170,100)(210,100){4}{6}
\ArrowLine(170,40)(210,40)
\ArrowLine(210,100)(260,100)
\Text(222,70)[]{$\chi^-$}
\DashArrowLine(210,40)(260,40){4}
\ArrowLine(210,40)(210,100)
\Text(268,90)[]{$\tilde{\chi}^-$}
\Text(268,50)[]{$\tilde{\nu}_e$}
\end{picture}
\vspace*{-1.5cm}
\caption[]{Feynman diagrams contributing to the associated production
of selectrons with neutralinos (a) and sneutrinos with charginos (b) in 
$e^-\gamma$ collisions.} 
\vspace*{-.2cm}
\end{figure}

The cross sections of the subprocesses for longitudinally polarized electrons 
and unpolarized photons are as follows:  
\beq
\hat{\sigma} (e^-_{L,R} \gamma \to \tilde{e}^-_{L,R}  \chi^0_i)
&=& \frac{\pi \alpha^2} {2 \hat{s}^3} |G_{L,R}^i|^2 \Bigg[ (\hat{s}-7 \mx^2 + 
7 \me^2 )\sqrt{\lambda} \non \\
&& \hspace*{-2.5cm}  -4(\me^2-\mx^2)(\hat{s}+\me^2 -\mx^2) \ln \left( 
\frac{\hat{s}-\mx^2+\me^2+\sqrt{\lambda}}{\hat{s}-\mx^2+\me^2-\sqrt{\lambda}} 
\right) \Bigg] \label{neut} \\
\hat{\sigma}(e^-_{L} \gamma\to \tilde{\nu}_L \chi_i^-) 
&=& \frac{\pi \alpha^2} {2 \hat{s}^3}|G^i|^2  
\Bigg[ (-3\hat{s}-7\mxc^2+7\mn^2)\sqrt{\lambda} \non \\
&&  \hspace*{-2.5cm} + 2(\hat{s}^2+2\mxc^2\hat{s}
-2\mn^2\hat{s}+2 (\mxc^2-\mn^2)^2) 
\ln\left( \frac{\hat{s}+\mxc^2-\mn^2+\sqrt{\lambda}}{\hat{s}+\mxc^2
-\mn^2-\sqrt{\lambda}} \right) \Bigg] \label{char} 
\eeq
where $\hat{s}$ is the $e\gamma$ c.m.~energy and $\lambda$, the two--body phase 
space function, given by
\beq
\lambda \equiv \lambda (\hat{s}, m_{\tilde{\ell}}^2, m_{\chi_i}^2) =  
\hat{s}^2 + \ml^4 + \mxx^4 -2\hat{s}\ml^2 -2\hat{s}\mxx^2 -2\ml^2\mxx^2 
\label{lam}
\eeq
$G_{L,R}^i$ ($G^i$) denote the selectron--neutralino--electron 
(sneutrino--chargino--electron) couplings; in terms of
$s_W^2= 1-c_W^2 \equiv \sin^2\theta_W$ and  for $i=1,2$, they are given by:
\beq
G_L^i = \frac{1}{\sqrt{2}} \bigg( \frac{ N_{i1}}{c_W}+ \frac{N_{i2}} {s_W} 
\bigg) \ , \ 
G_R^i =- \frac{ \sqrt{2}}{c_W}  N_{i1} \ , \ 
G^i = - \frac{1}{s_W} V_{i1} \label{coup}
\eeq
where $N$ is the matrix which diagonalizes the $4\times 4$ neutralino matrix
while $V$ is one of the unitary matrices which diagonalize the $2\times 2$ 
chargino mass matrix \cite{Haber,PO}. \s

The cross sections of the full processes $e^+ e^- \to e^+ \tilde{e}^-_{L,R}
\tilde{\chi}^0_i$ and $ e^+ e^- \to e^+ \tilde{\nu}_L \tilde\chi_i^-$ are
obtained by folding the cross sections of the subprocesses eqs.~(\ref{neut})
and (\ref{char}), respectively, with the photon luminosity $P_{\gamma/e}(y)$,
in which $s$ denotes the total $e^+ e^-$ c.m.~energy and $y$ the fraction of 
the electron energy carried by the photon:
\beq
\sigma (e^+ e^- \to e^+ \tilde{\ell} \chi_i) = \int_{y^-}^{y^+} dy\;
P_{\gamma/e}(y) \; \hat{\sigma} (\hat{s} = y s)
\eeq

\nn In our analysis, we will discuss three possibilities for the photon 
luminosity: \bigskip

(a) The Weizs\"acker-Williams (WW) spectrum \cite{WW}
\beq
P_{\gamma/e}(y) = \frac{\alpha}{2\pi}\frac{1+(1-y)^2}{y} 
\ln\frac{s}{m_e^2}
\eeq
with the electron mass $m_e=510.99906$~keV and the integration boundaries
\beq
y^- = (m_{\tilde{\ell}}^2+m_{\chi_i}^2) /s \;, \quad y^+ = 1
\label{bound}
\eeq
In general, this approximation overestimates the production cross sections. 
\bigskip 

(b) The improved Weizs\"acker--Williams approximation \cite{IWW} is more
reliable than the original WW approximation for small enough $\theta_c$ values,
where $\theta_c$ denotes the angle between the direction of the positron before
and after radiating off a photon [outgoing positrons above the angular cut
$\theta_c$ compared to the initial direction of the beam are eliminated]. 
Denoting the electron (beam) energy by $E=\sqrt{s}/2$, the spectrum is given by
\beq
P_{\gamma/e}(y) &=& \frac{\alpha}{2\pi} \Bigg[ 2(1-y) \bigg( \frac{m_e^2 y}
{E^2(1-y)^2 \theta_c^2 + m_e^2 y^2} - \frac{1}{y} \bigg) \non \\
&& \ \ + \frac{1+(1-y)^2}{y} \ln \frac{E^2(1-y)^2\theta_c^2+m_e^2 y^2}{m_e^2 y^2} 
+ {\cal O} (\theta_c,m_e^2/E^2) \Bigg]
\eeq

(c) The Compton back--scattering spectrum \cite{laser} 
\beq
P_{\gamma/e}(y) = \frac{1}{\sigma_c} \frac{d\sigma_c (y)}{dy}
\eeq
with the differential Compton cross section 
\beq
\frac{d\sigma_c}{dy}  = \frac{\pi\alpha^2}{x m_e^2} [f_0 + P_e P_\gamma f_1 
+ P_e P_{\gamma '} f_2 + P_\gamma P_{\gamma '} f_3]
\eeq
$P_e,P_\gamma,P_{\gamma '}$ denote the helicities of the initial electron, 
the laser photon $\gamma$ and the scattered photon $\gamma '$, respectively, 
with $-1 \le P_e,P_\gamma,P_{\gamma '} \le 1$. The parameter
$x$ is given by $x = 4 E \omega_0/m_e^2$ with $E=\sqrt{s}/2$ and the 
laser energy is $\omega_0 = 1.26$~eV. In terms of $r=y/(x(1-y))$, the functions 
$f_i$ (i=0,...,3) read
\beq
f_0 = \frac{1}{1-y} + 1-y - 4r(1-r) \;, \quad f_1 = xr (1-2r)(2-y) \non\\
f_2 = xr(1+(1-y)(1-2r)^2) \;, \quad f_3 = (1-2r)(\frac{1}{1-y}+1-y) 
\eeq
The integrated Compton cross section $\sigma_c$ can be cast into the form 
\beq
&& \hspace*{1cm} \sigma_c = \sigma_c^{\rm np} + P_e P_\gamma \sigma_c^{\rm p}
\non \\ 
\sigma_c^{\rm np} &=& \frac{\pi\alpha^2}{x m_e^2} \Bigg[ \frac{1}{2} + 
\frac{8}{x} -\frac{1}{2(1+x)^2} + \bigg( 1-\frac{4}{x}-\frac{8}{x^2} 
\bigg) \ln (x+1) \Bigg] \non \\
\sigma_c^{\rm p} &=& \frac{\pi\alpha^2}{x m_e^2} \Bigg[ -\frac{5}{2} + 
\frac{1}{x+1} - \frac{1}{2(x+1)^2} + \bigg(1+\frac{2}{x}\bigg) 
\ln (x+1) \Bigg]
\eeq
with the integration boundaries in this case,  given by 
\beq
y^- = (m_{\tilde{\ell}}^2+m_{\chi_i}^2) /s \;, \quad 
y^+ = x/(x+1)
\eeq
In $e\gamma$ colliders, with the photon generated by Compton
back--scattering of laser light, c.m.\ energies of the order of 80 to 90\% 
of the $\ee$ collider energy, integrated luminosities $\int {\cal L} \sim 
200$ fb$^{-1}$, and a high degree of longitudinal photon polarization can be 
reached \cite{newgamma}. \s

The production cross sections of selectrons and sneutrinos in association with,
respectively, the two lightest neutralinos $\chi_{1,2}^0$ and the lighter
chargino $\chi_1^\pm$ are shown as functions of the slepton masses in Figs.~2
for the three options of the photon luminosity discussed above:  2a for the WW
approximation, 2b for the improved WW approximation with an angular cut
$\theta_c=1^\circ$ and 2c for the Compton collider. The $\ee$ c.m.~energies are
set to 500 GeV (left panel) and 1 TeV (right panel) as expected for the next
generation colliders \cite{NLC}. The initial $e^-$ is assumed to be polarized
longitudinally: left--handed polarization for the production of $\tilde{e}_L$
and $\tilde{\nu}_e$ and right--handed polarization for the production of
$\tilde{e}_R$. The degrees of polarization are assumed to be equal to one,
close to the value expected ($80$ to $90\%$) at machines such as TESLA 
\cite{NLC}. Note that only processes involving the $e^-\gamma$ initial state
are taken into account in the cross sections of Figs.~2a--b; inclusion of the 
charge conjugate processes with $e^+ \gamma$, would lead to cross sections 
which are larger by a factor of two [if the degree of polarization of the 
$e^+$ beam is the same as for the $e^-$ beam]. \s 

For the SUSY parameters, we have chosen $\tb=30$, the higgsino mass parameter
$\mu=500$ GeV and the gaugino masses at the weak scale $M_1=50$ and
$M_2=100$ GeV, which implies approximate unification at the GUT scale. This
leads to gaugino--like lighter charginos and neutralinos with masses
$m_{\chi_1^0}\simeq M_1$ and $m_{\chi_2^0} \simeq m_{\chi_1^\pm} \simeq M_2$, 
i.e. around the experimental limits from LEP2 negative searches \cite{LEP2}. 
This choice is motivated by the fact that in this limit, the
selectron--electron--$\chi^0_{1,2}$ and sneutrino--electron--$\chi_1^-$
couplings eq.~(\ref{coup}) are maximal. For mixed gaugino and higgsino states,
i.e. for $|\mu| \sim M_2$, these couplings are suppressed and vanish in the
higgsino--like limit $|\mu| \ll M_2$ [the higgsino couplings to electrons and
sleptons are proportional to the electron mass]. We will not discuss the case
of associated production of the heavier chargino and neutralino states with
sleptons which are disfavored by phase--space. In particular, for the case of
interest, i.e. for $M_2 \ll |\mu|$, the heavier chargino and neutralino states
are higgsino--like with large masses $m_{\chi_3^0} \sim m_{\chi_4^0} \sim
m_{\chi_2^\pm} \simeq |\mu|$ and in addition, they have tiny couplings to
electron--slepton pairs.  For more details, see the discussion given in
Ref.~\cite{first}. Note also that the cross section for the process $e^-
\gamma \to \tilde{e}_R^- \chi_2^0$ vanishes in the limit of gaugino--like
next--to--lightest neutralinos since the right--handed selectron does not
couple to the winos. \s

\begin{figure}[htbp]
\begin{center}
\vskip-5.5cm
\mbox{\hskip-3cm\centerline{\epsfig{file=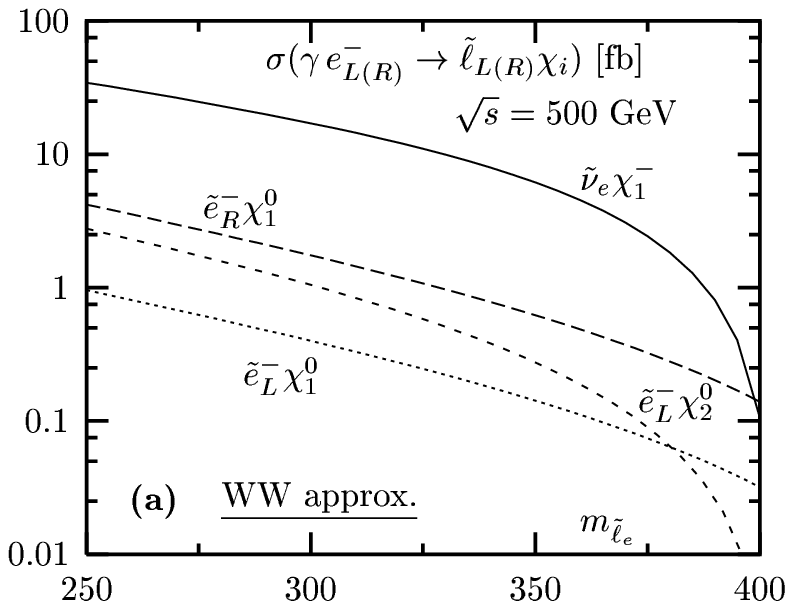,width=21cm}} 
\hskip-7.9cm\centerline{\epsfig{file=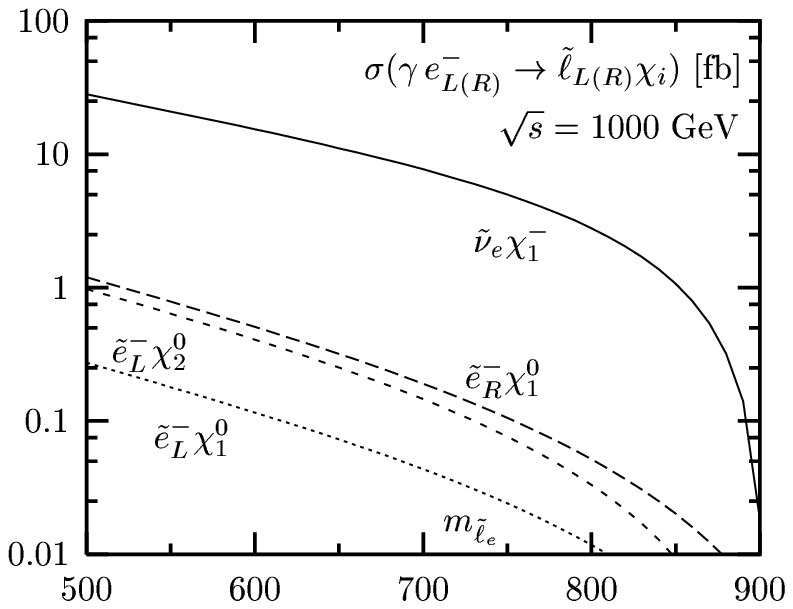,width=21cm}} }
\vskip-23.cm
\mbox{\hskip-3cm\centerline{\epsfig{file=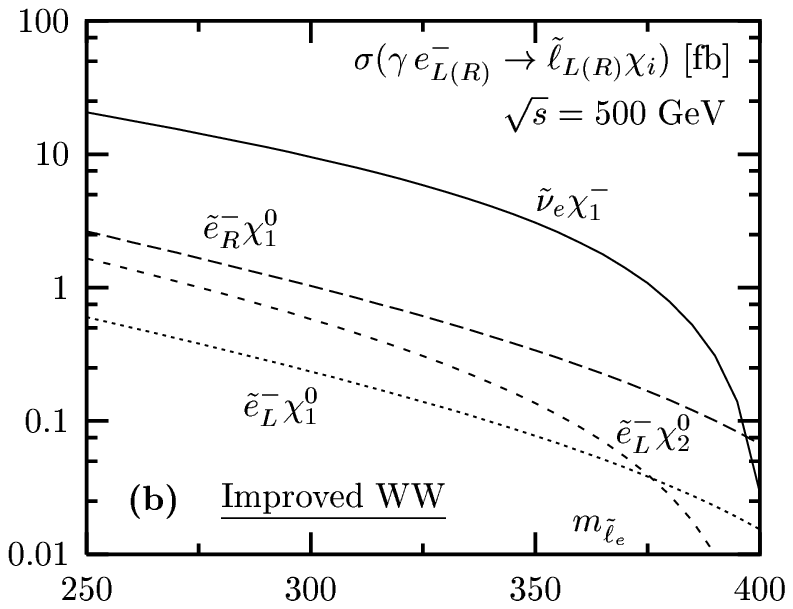,width=21cm}} 
\hskip-7.9cm\centerline{\epsfig{file=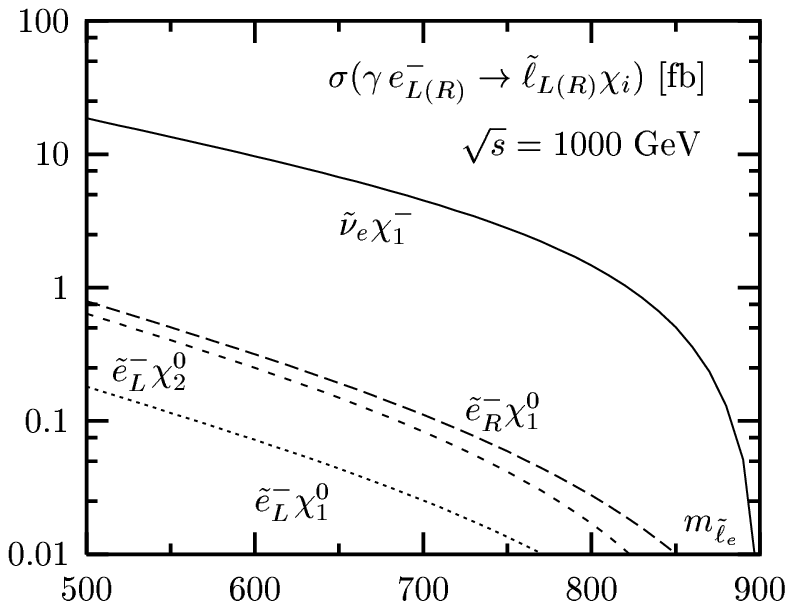,width=21cm}} }
\vskip-23.cm
\mbox{\hskip-3cm\centerline{\epsfig{file=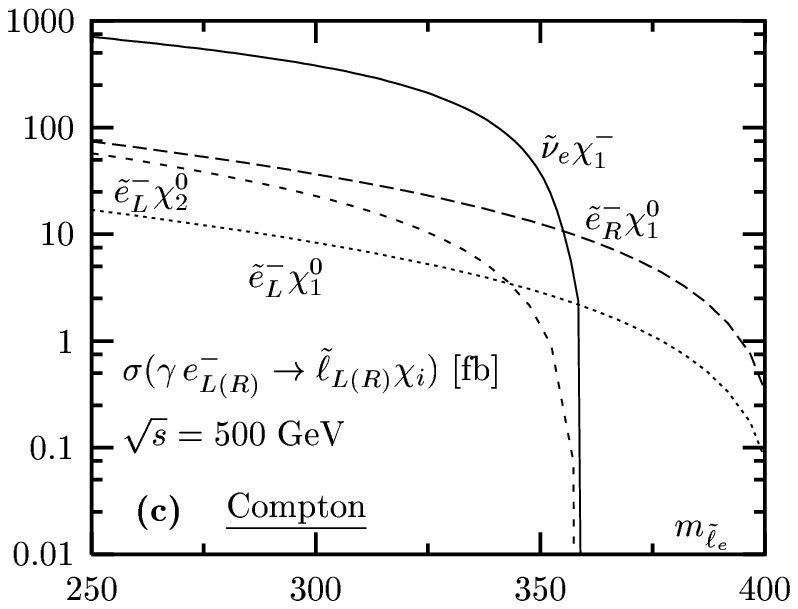,width=21cm}} 
\hskip-7.9cm\centerline{\epsfig{file=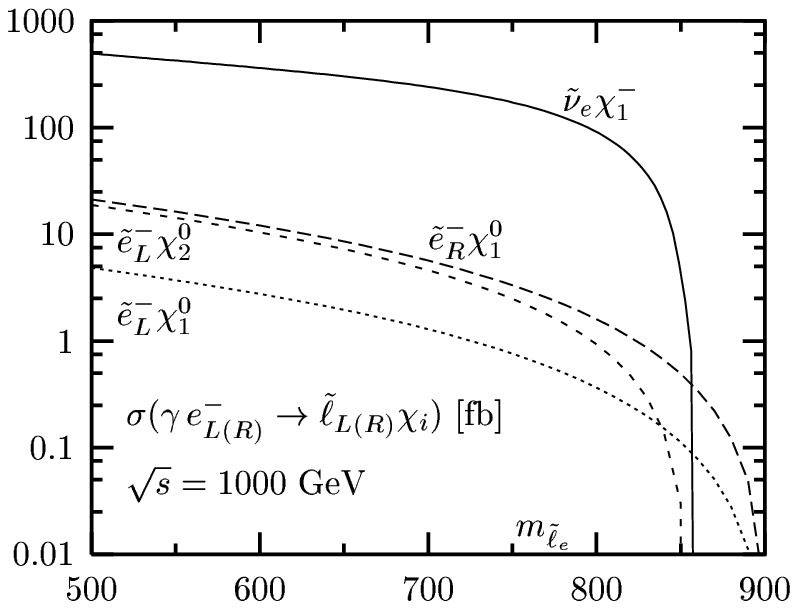,width=21cm}} }
\end{center}
\vspace*{-19cm}
\caption{Total cross sections for the associated production  of first 
generation sleptons with charginos and neutralinos in $e^-\gamma$ collisions
as functions of the slepton masses for c.m. energies of 500 GeV (left panel) 
and 1 TeV (right panel) in the WW approximation, improved WW approximation 
and for a Compton collider.} 
\end{figure}

For the $\ee$ option with Weizs\"acker--Williams photons, Fig.~2a and 2b, the
cross sections in the original approximation are about a factor of two larger
than in the improved approximation with the angle $\theta_c$ between the
initial and the final positron set to $1^\circ$. The cross section in the
improved WW approximation vanishes at $\theta_c=0$ [since the photon luminosity
in eq.~(10) goes to zero] and increases with increasing $\theta_c$ [some
illustrative examples will be given later on].  In the chosen scenario, for
slepton masses only slightly above the threshold for pair production in $\ee$
collisions, $\sqrt{s}\sim 2m_{\tilde{\ell}}$, and fairly far from the
kinematical limit in associated production, $\sqrt{s} \sim m_{\tilde{\ell}} +
m_\chi$, the gaugino mass effects do not play a significant role and the cross
sections for selectron production are such that 
\beq
\sigma(e_L \gamma \to \tilde{e}_L \chi_2^0)/ \sigma(e_L \gamma \to
\tilde{e}_L \chi_1^0) \sim 3 \ \ , \  \ \sigma(e_R \gamma \to \tilde{e}_R 
\chi_1^0)/ \sigma(e_L \gamma \to \tilde{e}_L \chi_1^0) \sim 4  
\eeq
These are simply the ratios of the squares of the selectron--electron--gaugino
couplings in the gaugino--limit, $(G_L^2)^2/(G_L^1) ^2\sim 3$ (for $s_W^2
\simeq 1/4$) and $(G^1_R)^2/(G_L^1)^2=4$. The behaviour of the cross sections
are almost identical except for the more rapid fall--off for higher masses for
$\sigma(e_L \gamma \to \tilde{e}_L \chi_2^0)$, because of a fast reducing
phase--space [$m_{\chi_2^0}\simeq 2 m_{\chi_1^0}$]. \s

At a 500 GeV $\ee$ collider, the cross sections are of the order of 0.6 (2.6)
fb for $e_{L(R)} \gamma \to \tilde{e}_{L(R)} \chi_1^0$ for selectron masses of
$m_{\tilde{\ell}_{L,R}} \sim 275$ GeV in the WW approximation. This means that
300 (1300) events can be collected for a yearly integrated luminosity of $ \int
{\cal L}= 500$ fb$^{-1}$ as expected for a machine like TESLA. Even for masses
of about $m_{\tilde{\ell}} \sim 400$ GeV, i.e. far above the mass reach for
slepton pair production, the cross section for $\tilde{e}_R \chi_1^0$
production is of the order of 0.1 fb which means that approximately 100 events
can be collected in two years of running, a sample which might be sufficient to
discover these particles in the clean environment of $\ee$ colliders. The cross
sections will of course be smaller for increasing values of the gaugino masses 
as will be illustrated later.\s

The cross section for sneutrino--chargino production is an order of magnitude
larger than the largest cross section for selectrons, $\sigma(e_R \gamma \to
\tilde{e}_R \chi_1^0)$; for low slepton masses, it exceeds the level of 10 fb
(for $m_{\tilde{\nu}}=300$ GeV) at $\sqrt{s}=500$ GeV. This is mainly due to the
fact that the magnitude of charged currents is larger than the one for neutral
currents, but also because in the $t$--channel diagrams, it is a lighter
chargino which is exchanged in this case [compared to the slepton in the
previous case]. The latter feature also explains the slightly different
behaviours of the cross sections for higher slepton masses. \s

The cross sections are also shown for an $\ee$ c.m. energy of 1 TeV and slepton
masses above 500 GeV. While in the case of selectrons they are smaller by a
factor 3 to 4 [the $s$--channel electron exchange contribution scales like
$1/s$ and is therefore smaller] for a fixed $m_{\tilde{\ell}}/\sqrt{s}$ value, 
the cross sections stay approximately the same for sneutrino--chargino 
production because of the dominance of the chargino $t$--channel exchange. \s

Finally, the cross sections in the $e\gamma$ mode of the $\ee$ collider, with
the photon generated by Compton back--scattering, are shown in Fig.~2c.  The
laser energy is taken to be 1.26 eV, leading to a parameter $x=4.83$ and
$x=9.65$ for the c.m. energies of the initial $\ee$ machine of 500 GeV and 1
TeV, respectively. For relatively low slepton masses, the cross sections are
more than one order of magnitude larger than in the Weizs\"acker--Williams
approximation, with the ratios between cross sections with different states
being approximately the same. This means that if the luminosity of the
$e\gamma$ collider is of the same order as the luminosity of the $\ee$
colliders, a very large number of events can be collected. However, there 
is a very strong fall--off of some of the cross sections for large slepton 
masses, in particular in the case of selectron and sneutrino production
in association with $\chi_2^0$ and $\chi_1^\pm$, respectively. Because the
energy of the $e\gamma$ collider peaks at approximately 90\% of the $\ee$ c.m. 
energy, the phase space suppression becomes very strong for slepton masses
close to $m_{\ell} \sim 350$ GeV at $\sqrt{s}_{\ee}= 500$ GeV, since in this
case, $m_{\chi_2^0} \sim m_{\chi_1^\pm} \sim M_2 =100$ GeV leading to $m_{\ell}
+ m_{\chi_i} \sim 450$ GeV. \s

Thus the cross sections for associated production of sleptons of the first
generation along with the lighter charginos or neutralinos are significant,
even in the $\ee$ option with  Weizs\"acker--Williams photons. Selectrons or
sneutrinos with masses beyond the values which can be reached in pair
production in $\ee$ collisions [approximately the beam energy] can be probed
with high luminosities if the charginos and neutralinos are lighter. However,
the accompanying electron/positron are in general not experimentally detected
in this case. To take into account this possibility as well, the full
three--body production process, $\ee \to \ell \tilde{\ell} \chi_i$, to which we
turn now, has to be considered.  

\subsection*{3. Production in $\ee$ collisions} 

There is a large number of Feynman diagrams contributing to the associated
production in $\ee$ collisions of first generation selectrons $\tilde{e}_{L,R}$
or sneutrinos $\tilde{\nu}_{e}$ with the lighter chargino $\chi_{1}^\pm$ or
neutralinos $\chi_{1,2}^0$. Neglecting the exchange of the heavier chargino or
neutralinos [which, as discussed previously, do not couple to the sleptons and
leptons in the gaugino limit] one has: 28 Feynman diagrams for $\ee \to e^+
\tilde{e}_{L,R}^- \chi_i^0$, 17 diagrams for $\ee \to \bar{\nu}_e \tilde{\nu}_e
\chi_i^0$, 13 diagrams for $\ee \to \nu_e \tilde{e}_{L}^- \chi_1^+$, 6 diagrams
for $\ee \to \nu_e \tilde{e}_{R}^- \chi_1^+$ and 12 diagrams for $\ee \to e^+
\tilde{\nu}_{e} \chi_1^-$.  The set of generic diagrams contributing to the
associated production of left-- or right--handed selectrons with the
neutralinos $\chi_{1,2}^0$ are shown in Figs.~3a--c. A similar set of diagrams
appears in the other processes. These diagrams can be divided into three
separately gauge invariant categories: \s

({\bf a}) Universal diagrams which occur also in the case of the associated 
production of second and third generation sleptons with electroweak gauginos
[and as such, this set is therefore gauge invariant by itself]. 
The amplitudes are discussed in Ref.~\cite{first}. \s

({\bf b}) Non--universal diagrams, i.e. which are special to the case of first
generation sleptons, and which do not involve the exchange of photons in the
$t$--channel. They all involve neutralino or $Z$ boson exchange in the
$t$--channel [which does not occur for associated smuon production due to
flavor conservation]. \s

({\bf c}) Non--universal diagrams where a photon is exchanged in the
$t$--channel. Here, the electron or positron radiates a photon and part of the
reaction will be exactly the same as the one discussed previously for $e
\gamma$ collisions, but with the photon being virtual.  \s

Similar diagrams occur for the other associated production processes. 
Note that in the processes where charginos and sneutrinos are produced in
association, $\ee \to e^\pm \tilde{\nu}_e \chi_1^\mp$, there are also Feynman
diagrams involving the  $t$--channel exchange of photons, Fig.~3d, and
which correspond to those discussed in the case of $e\gamma$ collisions but with
the photon being virtual. Such diagrams with $t$--channel photons do not occur 
in the associated production of charginos with selectrons or neutralinos with
sneutrinos.  \s

The amplitudes for diagrams 3b, 3c and 3d are given in Appendix A, while the
ones for diagrams 3a are already given in Ref.~\cite{first} from which we
borrow the notation. The full analytical expression for the differential cross
section is rather lengthy and not very telling. In Appendix B, we will however
write down the expressions for the squared amplitude of the sum of the two
contributions which involve the $t$--channel photon exchange only. This, as we
will show later, is the most important one numerically.  In the following, we
will simply discuss our numerical results from the complete calculation. We
note that in all cases, we have thoroughly cross checked our numerical results
against the corresponding ones from the package {\tt CompHEP} \cite{Comphep}.

\begin{figure}[htbp]
\vspace*{-2.7cm}
\centerline{\epsfig{file=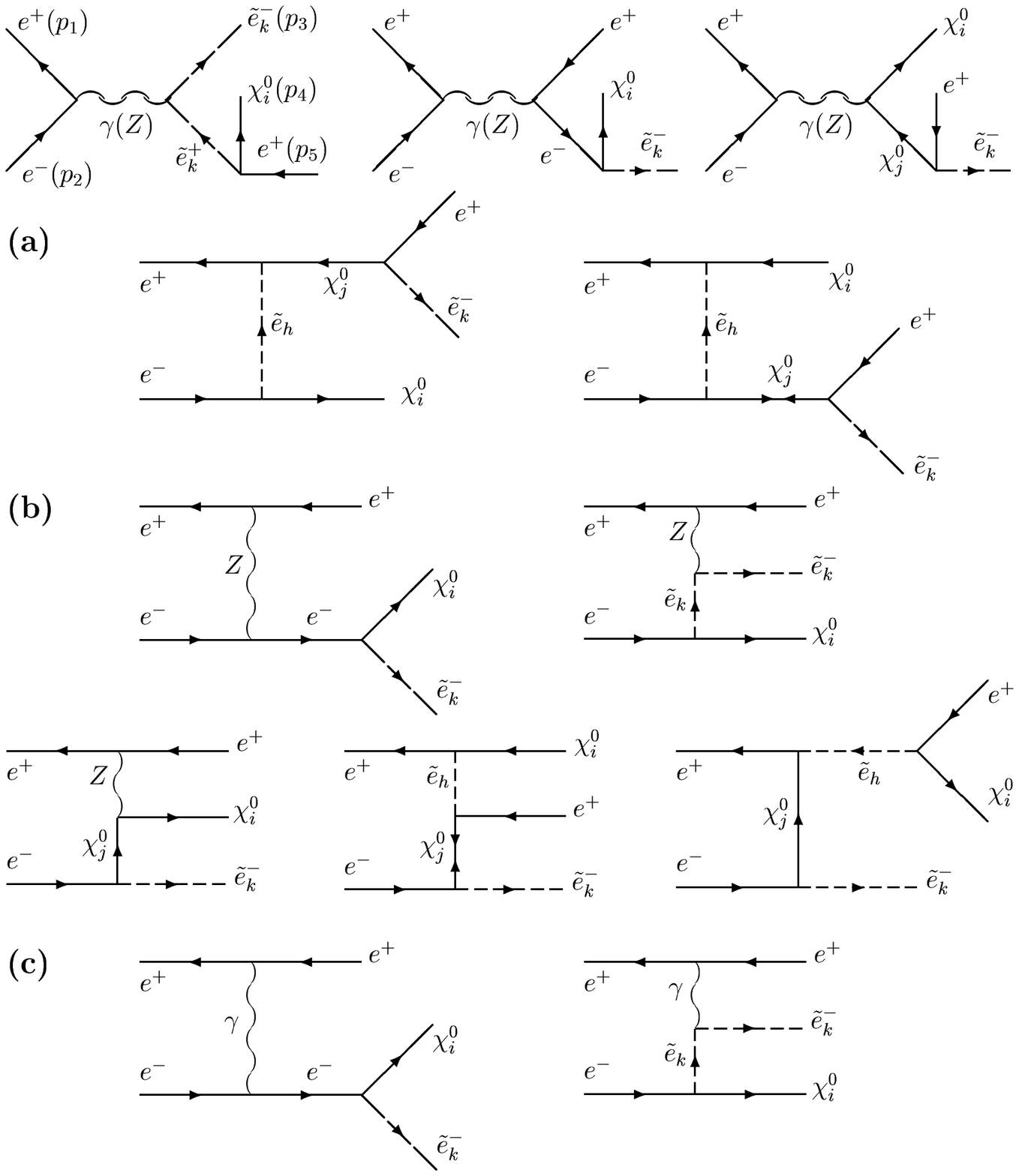,width=18cm}}
\vskip-10cm
\centerline{\epsfig{file=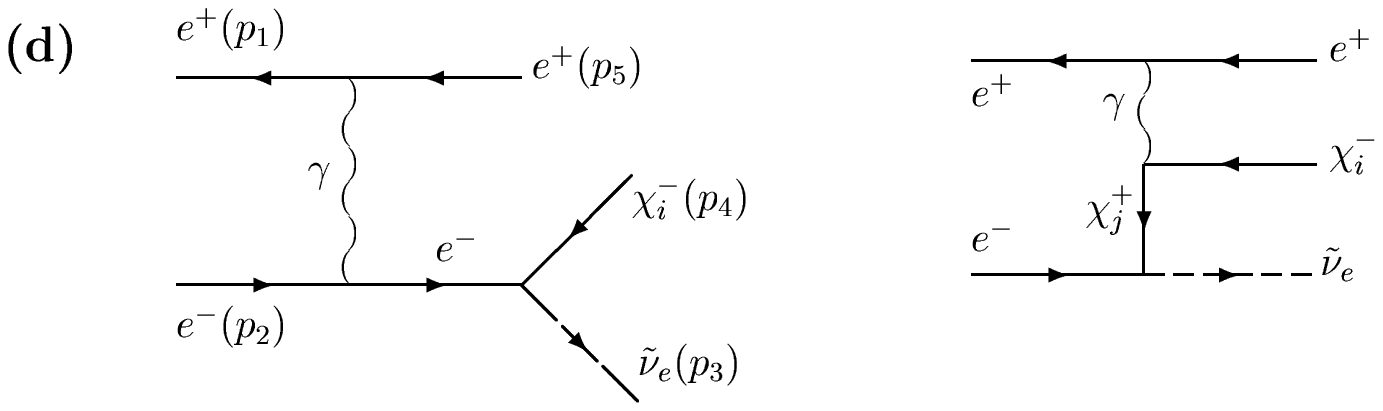,width=18cm}}
\vskip-20cm
\caption{(a--c): Feynman diagrams for the associated production of selectrons 
with neutralinos, $\ee \to e^+ \tilde{e}_{L,R}^- \chi_i^0$: (a) universal 
diagrams, (b) non--universal $t$--channel diagrams without $\gamma$ exchange 
and (c) diagrams with $t$--channel $\gamma$ exchange; (d) is for diagrams with 
$t$--channel photon exchange for associated sneutrino--chargino production.}
\end{figure}

\newpage

The total cross sections for the associated productions of first generation
sleptons with  the two lightest neutralinos $\chi_1^0$ (a) and $\chi_2^0$ (b)
as well as with the lightest chargino $\chi_1^\pm$ (c) are shown in Fig.~4 as
functions of the slepton mass and for two center of mass energies $\sqrt{s}
=500$ GeV (left panel) and 1 TeV (right panel). As in the previous section, we
have assumed a common soft--SUSY breaking mass for the scalars and fixed the
SU(2) gaugino and higgsino mass terms to be $M_2=2M_1=100$ GeV and $\mu=500$
GeV with $\tb=30$. The cross sections are displayed for unpolarized initial
beams and the charged conjugate final states [which lead to  a factor of two
increase of the cross sections] have been included. In the case of $e^\pm
\tilde{e}_{L,R}^\mp \chi_{1,2}^0$ and $e^\pm \tilde{\nu}_e \chi_1^\mp$
production, which involve the $t$--channel photon poles, we have applied a cut
on the minimum angle between the direction of the final electron (positron)
with respect to the incident electron (positron), $\theta_{\rm min}=1^\circ$. 
This cut is used to avoid the numerical instabilities near the $t$--channel
photon pole\footnote{In principle, one can regulate this pole by including the
finite mass of the electron also in the phase space (as we have already done).
However, since $m_e$ is much smaller compared to $\sqrt{s}$, a very delicate
and lengthy Monte--Carlo phase--space integration would be required.}.  \s

As can be seen, at $\sqrt{s}=500$ GeV, the production cross sections can be
rather large, exceeding the level of 0.1 fb for slepton masses close to
$m_{\tilde{\ell}} \sim 350$ GeV, except in the case of $\tilde{e}_R$ production
with $\chi_2^0$ and $\chi_1^+$ [because $\tilde{e}_R$ has very small couplings
to these dominantly SU(2) gauginos] and $\tilde{\nu}_e$ production with the LSP
[because the bino-$\tilde{\nu}_e$-$\nu_e$ coupling is suppressed and there is
no $t$--channel photon to enhance the rate]. The largest rates are obtained for
the processes $\ee \to e^\pm \tilde{e}_R^\mp \chi_1^0$ and $\ee \to e^\pm
\tilde{\nu}_e^{(*)} \chi_1^\mp$. This is due to the larger $\tilde{e}_R$
couplings to electrons and binos in the former case and because of the stronger
charged current interactions for the latter, both of which are fortified with
contributions from the $t$--channel photon exchange. For $m_{\tilde{\ell}} \sim
350$ GeV and $m_{\chi_1^+} \sim 2m_{\chi_1^0} \sim 100$ GeV, the cross sections
are at the level of 0.5 fb and 2 fb, respectively, allowing one to collect a
few hundreds of events in each process with the expected luminosity of $\int
{\cal L} \sim 500$ fb$^{-1}$. [In fact, even for $m_{\tilde{\ell}} \sim 400$
GeV, one can still produce a few tens of events per year in this case.] At
$\sqrt{s}=1$ TeV, the cross sections are in general smaller than that at a 500
GeV collider for a given ratio $m_{\tilde{\ell}}/\sqrt{s}$ [with all the other
parameters fixed] except in some cases where sleptons are produced in
association with the heavier gaugino-like $\chi_2^0$ and $\chi_1^\pm$ for which
significantly larger regions of phase space are now available. \s

\begin{figure}[htbp]
\begin{center}
\vskip-5.5cm
\mbox{\hskip-3.5cm\centerline{\epsfig{file=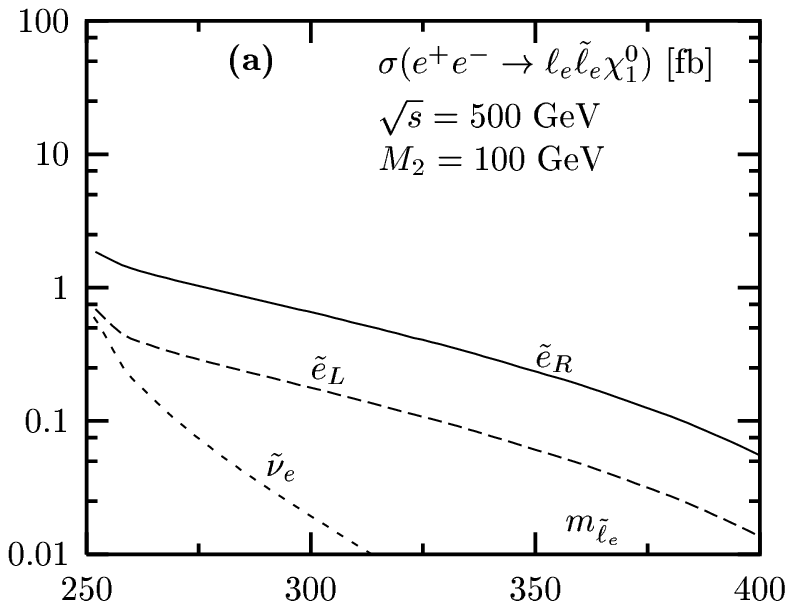,width=21cm}} 
\hskip-7.5cm\centerline{\epsfig{file=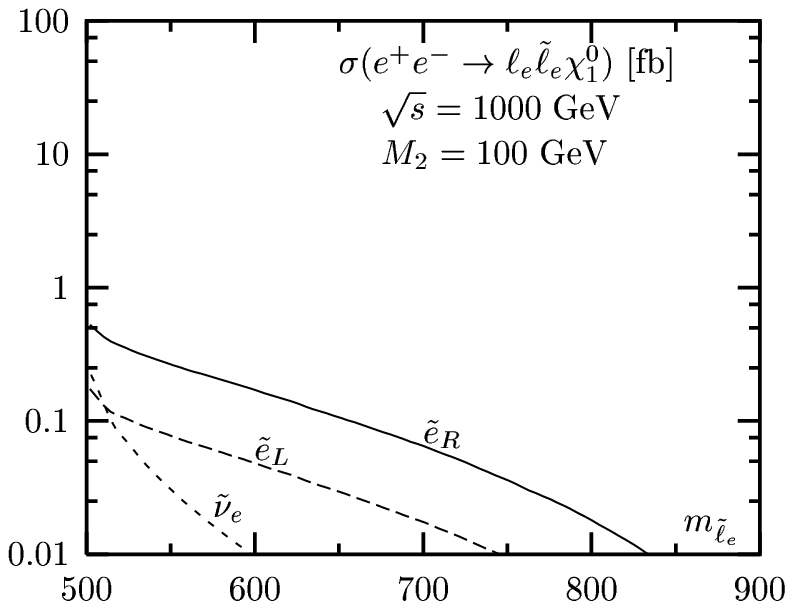,width=21cm}} }
\vskip-23.cm
\mbox{\hskip-3.5cm\centerline{\epsfig{file=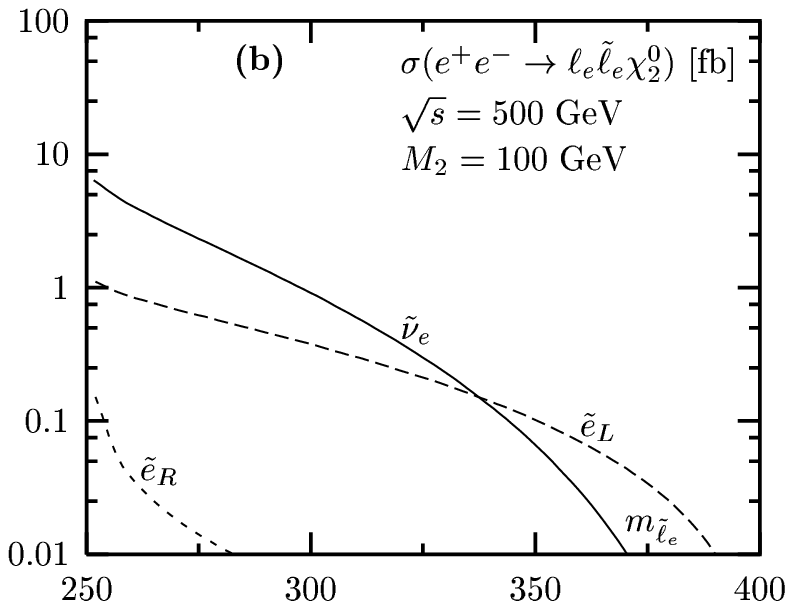,width=21cm}} 
\hskip-7.5cm\centerline{\epsfig{file=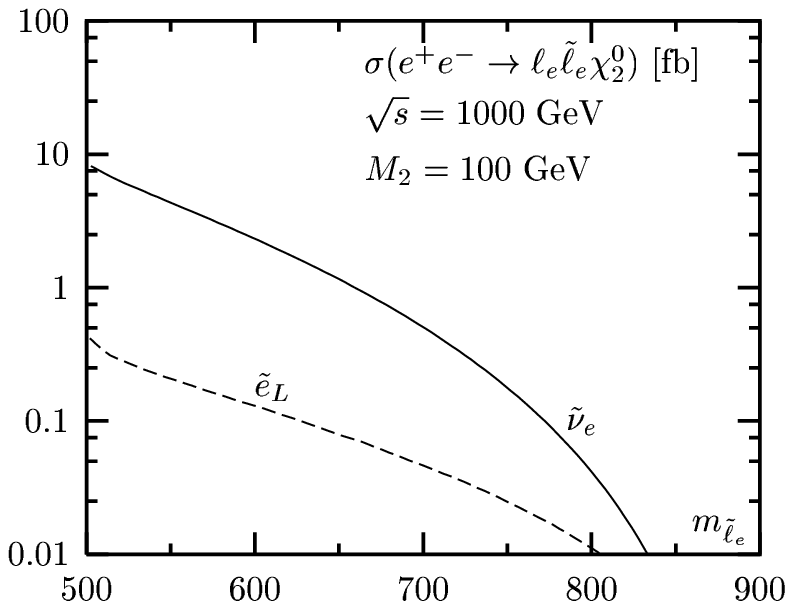,width=21cm}} }
\vskip-23.cm
\mbox{\hskip-3.5cm\centerline{\epsfig{file=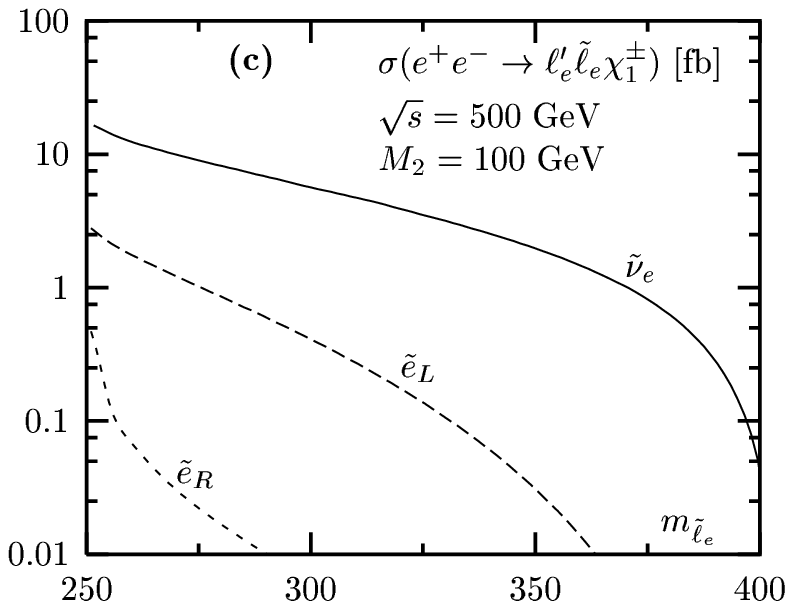,width=21cm}} 
\hskip-7.5cm\centerline{\epsfig{file=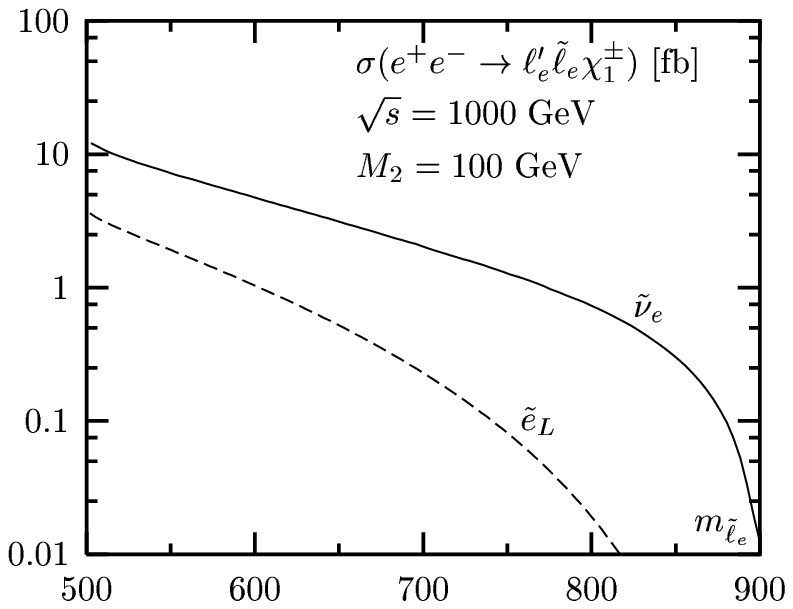,width=21cm}} }
\end{center}
\vspace*{-19cm}
\caption{The total cross sections for the associated production processes 
of first generation sleptons and electroweak gauginos as functions of the 
slepton masses for center of mass energies of 500 GeV (left panels) and 1 
TeV (right panels).} 
\end{figure}

In Fig.~5, we display the cross sections at $\sqrt{s}=500$ GeV for a fixed
$m_{\tilde{\ell}}\sim 275$ GeV, as functions of the gaugino mass parameter $M_2$, i.e.~for somewhat heavier $\chi_1^\pm$ and $\chi_{1,2}^0$ gaugino states.  As
expected, the cross sections decrease with larger  $M_2$, in particular for
associated slepton productions with heavier gauginos, $\chi_2^0$ and
$\chi_1^\pm$. In the case of $\tilde{e}_{L,R}$ production with the LSP, where
more phase space is available since $m_{\chi_1^0} \sim 0.5 M_2$, the cross
sections drop by less than a factor of 10 for a variation of $M_2$ from 100 to
250 GeV. \s 


\begin{figure}[htbp]
\begin{center}
\vskip-5.5cm
\mbox{\hskip-3.5cm\centerline{\epsfig{file=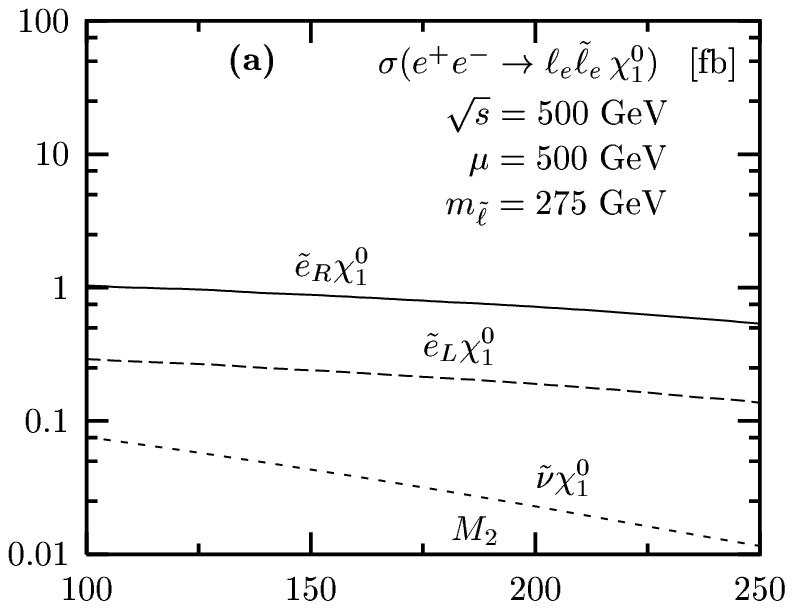,width=21cm}} 
\hskip-7.5cm\centerline{\epsfig{file=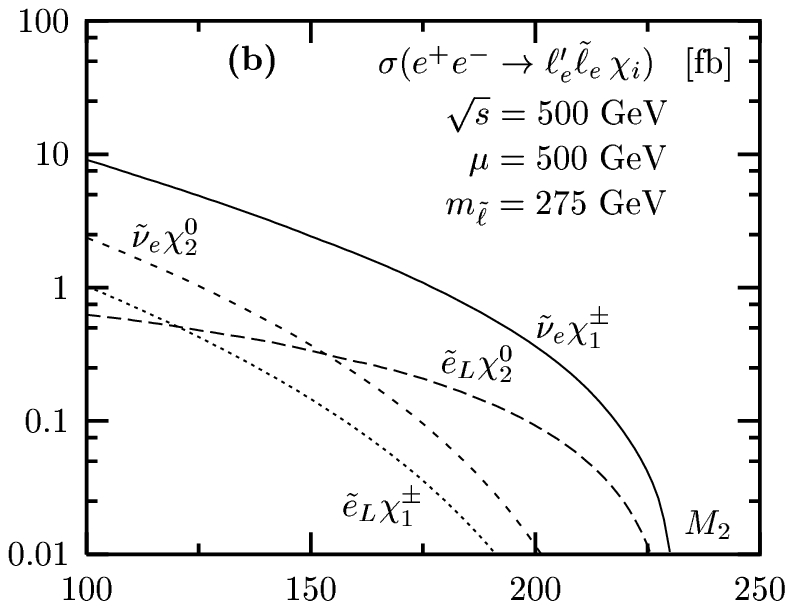,width=21cm}} }
\end{center}
\vspace*{-19.3cm}
\caption{The total cross sections for the associated production of first 
generation sleptons with electroweak gauginos as functions of $M_2$ for 
$\sqrt{s}=500$ GeV and $m_{\tilde{\ell}}= 275$ GeV.} 
\end{figure}

Finally, Fig.~6 shows the cross sections for the associated slepton 
production with the LSP at $\sqrt{s}=500$ GeV as functions of the slepton masses
when the lighter neutralinos and chargino are almost higgsino--like (a) and
mixtures of gauginos and higgsinos (b). For higgsino--like $\chi_1^0$ particles 
the cross sections are much smaller than the previous case, a result of the 
smaller lepton--slepton--LSP coupling. In the case of mixed $\chi_1^0$ states, 
the cross sections are slightly smaller than for a gaugino--like LSP. However, 
since all neutralinos and charginos will have comparable masses in this case, 
one has to consider the production with the heavier $\chi$ particles. At the 
end of the day, once all processes have been included, the sum of the cross 
sections for the various processes will be similar in magnitude compared to 
the case of gaugino--like lightest neutralinos. 


\begin{figure}[htbp]
\begin{center}
\vskip-5.6cm
\mbox{\hskip-3.5cm\centerline{\epsfig{file=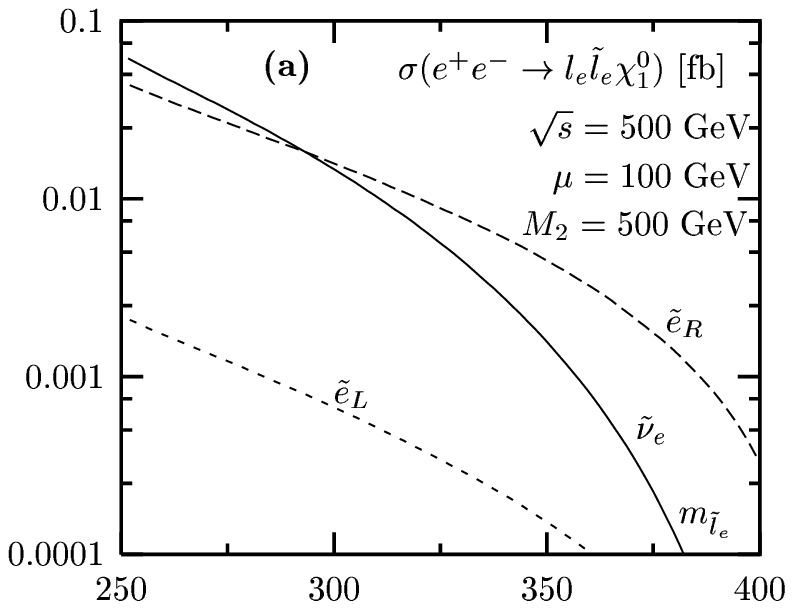,width=21cm}} 
\hskip-7.5cm\centerline{\epsfig{file=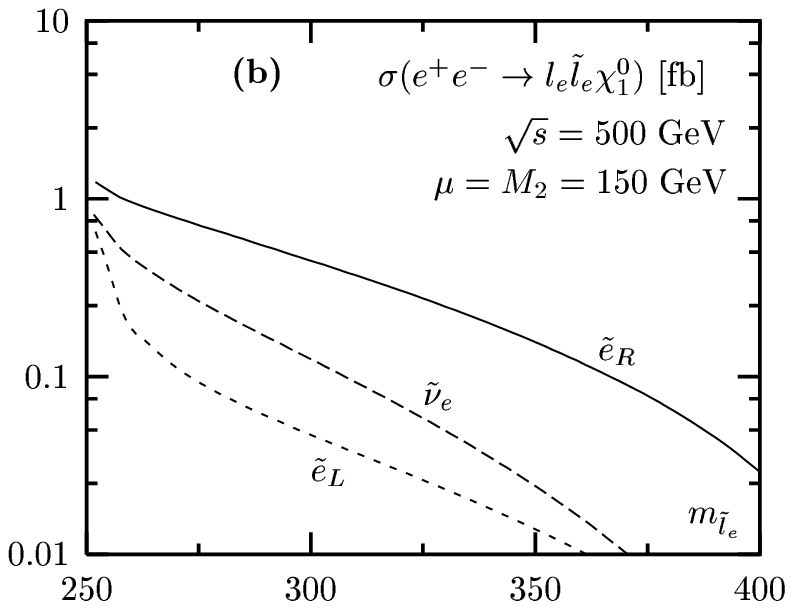,width=21cm}} }
\end{center}
\vspace*{-19.3cm}
\caption{The total cross sections for the associated production of first 
generation sleptons with the lightest neutralino as function of the slepton 
masses for $\sqrt{s}=500$ GeV for a higgsino--like $\chi_1^0$ (a) and a 
mixed higgsino--gaugino $\chi_1^0$ (b).} 
\end{figure}

\newpage

\subsection*{4. The total production cross sections}

Let us now discuss the magnitude of the various contributions to the total
associated cross section and the effects of the cut--off angle $\theta_{\rm
min}$ which in the previous discussion was set to $1^0$. Taking as examples the
associated production of the left-- and right--handed selectrons with the LSP,
$\ee \to e^\pm \tilde{e}_{L,R}^\mp \chi_1^0$, we display in Table 1 the cross
sections of the various contributions for a c.m.~energy $\sqrt{s}=500$ GeV and
the SUSY particle masses $m_{\tilde{e}_{L,R}}=275$ GeV and $m_{\chi_1^0}\sim 75$
GeV [$M_2=150$ GeV, $\mu=500$ GeV and $\tb=30$], for several values of the 
cut--off angle $\theta_{\rm min}$. We will also compare these with the results 
obtained in the improved Weizs\"acker--Williams approximation with a cut-off 
angle $\theta_c=\theta_{\rm min}$. 

\begin{table}[htbp]
\vspace*{-.5cm}
\renewcommand{\arraystretch}{1.2}
\begin{center}
$\tilde{e}_L$
\begin{tabular}{|c||c||c|c|c||c|} \hline
$\theta_c(\theta_{min})$ & IWW & Resonant & Non-resonant & Full & IWW+Full \\
\hline
\hline
$5^\circ$ & 0.363 & 0.125 & 0.023 & 0.148 & 0.511 \\
\hline
$4^\circ$ & 0.355 & 0.135 & 0.023 & 0.158 & 0.513 \\
\hline
$3^\circ$ & 0.343 & 0.146 & 0.023 & 0.169 & 0.512 \\
\hline
$2^\circ$ & 0.327 & 0.164 & 0.023 & 0.187 & 0.514 \\
\hline
$1^\circ$ & 0.300 & 0.195 & 0.023 & 0.218 & 0.518 \\
\hline
$0.5^\circ$ & 0.272 & 0.220 & 0.023 & 0.243 & 0.515 \\
\hline
$0.2^\circ$ & 0.236 & 0.269 & 0.023 & 0.292 & 0.528 \\
\hline
$0.1^\circ$ & 0.209 & 0.294 & 0.023 & 0.317 & 0.526 \\
\hline
$0.05^\circ$ & 0.182 & 0.317 & 0.023 & 0.340 & 0.522 \\
\hline 
\end{tabular}
\vskip 10pt
$\tilde{e}_R$
\begin{tabular}{|c||c||c|c|c||c|} \hline
$\theta_c(\theta_{min})$ & IWW & Resonant & Non-resonant & Full & IWW+Full \\
\hline
\hline
$5^\circ$ & 1.574 & 0.541 & -0.062 & 0.479 & 2.053 \\
\hline
$4^\circ$ & 1.535 & 0.580 & -0.062 & 0.518 & 2.053 \\
\hline
$3^\circ$ & 1.486 & 0.628 & -0.062 & 0.566 & 2.052 \\
\hline
$2^\circ$ & 1.417 & 0.712 & -0.062 & 0.650 & 2.067 \\
\hline
$1^\circ$ & 1.298 & 0.836 & -0.063 & 0.773 & 2.071 \\
\hline
$0.5^\circ$ & 1.180 & 0.963 & -0.063 & 0.900 & 2.080 \\
\hline
$0.2^\circ$ & 1.023 & 1.130 & -0.063 & 1.067 & 2.090 \\
\hline
$0.1^\circ$ & 0.905 & 1.272 & -0.063 & 1.209 & 2.114 \\
\hline
$0.05^\circ$ & 0.786 & 1.388 & -0.063 & 1.325 & 2.111 \\
\hline 
\end{tabular}
\end{center}
\noindent
\caption[]{Variation of the contributions to the production cross 
section (in femtobarns) for the final states $e^\pm \tilde{e}_L^\mp \chi_1^0$
(upper table) and  $e^\pm \tilde{e}_R^\mp \chi_1^0$ (lower table) with  the 
cut-off angle $\theta_{\rm min}$ and comparison with the improved WW 
approximation with $\theta_c = \theta_{\rm min}$. The parameters are 
$\sqrt{s}=500$ GeV, $m_{\tilde{e}_{L,R}}=275$ GeV, $M_2=150$ GeV,
$\mu=500$ GeV and $\tb=30$.}  
\end{table}

We define the ``resonant" contribution to be the cross section from the 
two diagrams involving the $t$--channel exchange of the photon in Fig.~3c. As
discussed previously, the sum of these two amplitudes is gauge invariant. These
contributions strongly depend on the value of the cut--off angle $\theta_{\rm
min}$ for the direction of the final electron (positron) with respect to the
initial electron (positron) beams and which vetoes final leptons going in the
very forward direction. The cross sections increase for decreasing cut--off
angles. For instance, they are a factor of 2 to 3 larger for $\theta_{\rm
min}\sim 0.5^\circ$ than for $\theta_{\rm min}\sim 5^\circ$, illustrating the
fact that in most cases the final lepton goes down in the forward direction.
However, for a small cut--off angle, we approach too closely the photon pole and
the phase--space integration becomes unstable numerically and when using a
Monte--Carlo method for such an integration, this actually calls for a very
large number of integration points and commensurate number of iterations to
achieve a stable result.  An intermediate value, say $\theta_{\rm min}\sim
1^\circ$ used in the previous discussion, is more appropriate since the cross
section is significant and at the same time its determination is more accurate.
\s

The ``non--resonant" contribution consists of the squared sum of the amplitudes
of the diagrams which do not involve the $t$--channel photon, Fig~(3a) and (3b)
[which forms also a gauge invariant set], and their interferences with the two
diagrams in Fig.~3c. This interference does not involve any pole [since there
is no $1/t^2$ terms] and is thus regular in the forward direction.  As can be
seen in Table 1, this contribution is rather small and does not depend strongly
on the cut--off angle $\theta_{\rm min}$ [provided that it is small enough]. 
The total cross section, i.e. the sum of the two contributions, is thus
dominated by the resonant piece\footnote{Note that the electron--photon vertex
in the resonant contribution should be evaluated at $q^2=0$ and therefore $
\alpha=1/137$ , while the other vertices should be evaluated at a scale close
to $\sqrt{s}$ and thus $\alpha \sim 1/128$ \cite{alpha}.  Also, in principle,
one should use a running $\alpha$; however, for our purpose, it is a good
approximation to use the two values above. We thank E. Boos and T. Ohl for a
discussion on this point.}.  
 
We have also displayed the $e^\pm \gamma \to \tilde{e}_L^\pm \chi_1^0$ cross
section in the improved  Weizs\"acker--Williams approximation [for unpolarized
beams and including the two charge conjugate processes] with a cut-off angle
$\theta_c= \theta_{\rm min}$. Here, the cross section decreases with decreasing
angle since the probability of having the lepton emitted from the initial beam
with a smaller angle is smaller, as is evident from eq.~(10). \s

In fact the ``full" cross section and the cross section in the improved WW
approximation are complementary since they deal with complementary regions of
the phase space for $\theta_c=\theta_{\rm min}$.  The sum of the two gives the
cross section for the associated production of sleptons and gauginos in the
full angular range whether the final lepton is in the forward direction [i.e. 
experimentally undetected] or not. This sum, given in the last column of the
table, is practically constant since it changes by less than $\sim 3\%$ between
$\theta= 5^\circ$ and $\theta= 0.05^\circ$, while both of its components change
by more than 100\%. This modest shift is probably due to the fact that, at
larger angles the improved WW approximation is rather poor, while at smaller
angles, the three--body cross section is numerically unstable.  A median value
of $\theta \sim 1^\circ$, which has been adopted here, should therefore give an
approximately good answer for the exclusive associated slepton+gaugino cross
section\footnote{This way of calculating the full cross section for multi--body
final states involving photons in the $t$--channel, i.e. by calculating
separately the cross section in the improved WW approximation in a narrow cone
around the beam axis and the full multi--body process for the rest (which
eliminates the numerical instabilities at small angles) and then performing 
the sum, has been discussed in the context of single $W$ production at $\ee$ 
colliders \cite{poles}. This allows for fast Monte--Carlo analyses.}.

\subsection*{4. Conclusions} 

In this paper, we have analyzed the associated production of first generation 
sleptons with neutralinos and charginos in the MSSM, at future high--energy 
$e^+e^-$ colliders, extending on the work done in Ref.~\cite{first}. We
have discussed both the two--body particle production in the $e\gamma$ mode and
the three--body particle production in the $\ee$ mode of the collider. In the
latter case, we have shown that the production cross sections, thanks to a
strong enhancement due to the presence of $t$--channel photon exchange
contributions, can be rather large even for slepton masses significantly
exceeding the kinematical two--body reach of the $\ee$ collider, i.e. 
$m_{\tilde{\ell}} > \sqrt{s}/2$. This is particularly the case for the
production of right--handed selectrons with the lightest neutralino and the
production of sneutrinos with the lightest chargino, where slepton masses a few
ten GeV beyond the beam energy can be probed if the electroweak gauginos are
relatively light.   \s

The final states discussed in this paper should be clear enough to be detected
in the clean environment of $\ee$ colliders. However, to asses more firmly this
possibility, a detailed study \cite{Peter} of the signal and the various 
Standard Model backgrounds as well as the SUSY processes leading to the same 
final states [such as $\chi_1^+ \chi_1^-$ and $\chi_1^0 \chi_2^0$ pair
production for which the cross sections can be large], has to be performed. In
particular, since in the kinematical regions where the cross sections are
large, a final state electron or positron goes in the direction of the beam
pipe, large and dangerous backgrounds such as single $W$ boson production or
$\gamma \gamma$ collisions should be studied in detail.  These studies are
beyond the scope of this preliminary analysis and will be performed in a near
future.  \bigskip

\nn {\bf Acknowledgments}: We thank Ayres Freitas and Peter Zerwas for 
clarifying discussions and E. Boos and A. Pukhov for their help in using {\tt 
CompHEP}. AKD is supported by a MNERT fellowship while AD and MM are supported 
by the Euro--GDR Supersym\'etrie and by the European Union under contract
HPRN-CT-2000-00149.  

\newpage

\subsection*{Appendix A: The amplitudes of the various contributions}
\setcounter{equation}{0}
\renewcommand{\theequation}{A\arabic{equation}}

In this Appendix, we present the amplitudes for the diagrams in Fig. 3b--3d
(the ones for the universal contributions Fig.~3a are given in
Ref.~\cite{first}).  The first two combined represent a generic set of extra
Feynman diagrams for final states involving left and right handed selectron
along with neutralinos (as compared to the set in Fig. 3a which is generic to
all scalar final states), i.e. $e^+ e^- \to \tilde{e}_{L,R}^-  e^+ \chi_i^0$.
The set of two diagrams in Fig. 3d are for the final state involving electronic
sneutrino along with the chargino, i.e.  $e^+ e^- \to \tilde{\nu}_e e^+
\chi_i^-$,  and are the dominant ones containing photon poles when the outgoing
positron is scattered in the very forward direction. In a covariant gauge, the
amplitudes are given by:
\beq
M_{b1} &=&
\frac{-e^3/(\stwn3 \ctwn2)}{\{(p_1-p_5)^2-m_Z^2 \} \, (p_1+p_2-p_5)^2} \;
[ \bar{v}_e(p_1) \, \gamma^\alpha \, (c_L^e P_L+c_R^e P_R) \, v_e(p_5) ]
\non \\ && \hskip 100pt \times \,
[\bar{u}_{\chi{_i}}(p_4) \; G^{Aik}_{b 1} \, (\pn1+\pn2-\pn5) \, \gamma_\alpha
\, (c_L^e P_L+c_R^e P_R) \, u_e(p_2)] 
\non \\ \non \\ \non
M_{b2} &=& \frac{e^3 \, a_{Z\tilde{e}_k\tilde{e}_k} / (\stwn3 \ctwn2)}
{\{(p_1-p_5)^2-m_Z^2 \} \, \{ (p_1-p_3-p_5)^2 - m_{\tilde{e}_k}^2 \}} 
\non \\ &&  \hskip 34pt \times \,
[ \bar{v}_e(p_1) \, (\pn1-\pn5-2\!\!\pn3) \, (c_L^e P_L+c_R^e P_R) \, v_e(p_5) ] \;
[\bar{u}_{\chi{_i}}(p_4) \; G^{Aik}_{b 2} \, u_e(p_2)] \non \\ \non \\
M_{b3} &=& \frac{e^3 / (\stwn3 \ctwn2)}
{\{(p_1-p_5)^2-m_Z^2 \} \, \{(p_2-p_3)^2 - \mgj^2\}} \;
[ \bar{v}_e(p_1) \, \gamma^\alpha \, (c_L^e P_L+c_R^e P_R) \, v_e(p_5) ]
\non \\ &&  \hskip 73pt \times \,
[\bar{u}_{\chi{_i}}(p_4) \; (O^{ij}_L P_L+O^{ij}_R P_R) \: \gamma_\alpha 
\, (\pn2-\pn3+\mgj) \, G^{Ajk}_{b 3} \, u_e(p_2)] \non \\ \non \\
M_{b4} &=& -\biggl(\frac{e^3}{\stwn3} \biggr)\,
\sum_h \frac{[\bar{v}_e(p_1) \, G^{Bih}_{b 4} \: v_{\chi_i^0}(p_4)] \; 
[\bar{v}_e(p_2) \, G^{Ajk}_{b 4} \, (\pn2-\pn3-\mgj) \,  G^{Ajh}_{b 4} \, v_e(p_5)]}
{\{(p_1-p_4)^2-m_{\tilde{e}_h}^2 \} \, \{(p_2-p_3)^2 - \mgj^2\}} \non \\ \non \\
M_{b5} &=& \biggl(\frac{e^3}{\stwn3} \biggr)\,
\sum_h \frac{\, [\bar{v}_e(p_1) \, G^{Bjh}_{b 5} \, (\pn2-\pn3+\mgj) \: G^{Ajk}_{b 5} \: 
u_e(p_2)] \; 
[\bar{u}_{\chi{_i}}(p_4) \, G^{Aih}_{b 5} \, v_e(p_5)]}
{\{(p_1+p_2-p_3)^2-m_{\tilde{e}_h}^2 \} \, \{(p_2-p_3)^2 - \mgj^2\}} \non\\ \\ \non\\ 
M_{c 1(d1)} &=& -\biggl(\frac{e^3}{\stw} \biggr)\,
\frac{[\bar{v}_e(p_1) \, \gamma^\alpha \, v_e(p_5)] \; 
[\bar{u}_{\chi{_i}}(p_4) \; G^{Aik}_{c 1(d 1)}
\; (\pn1+\pn2-\pn5) \, \gamma_\alpha \, u_e(p_2)]}
{(p_1-p_5)^2 \, (p_1+p_2-p_5)^2} \non \\ \non \\
M_{c2} &=& \biggl(\frac{e^3}{\stw} \biggr)\,
\frac{[\bar{v}_e(p_1) \, (\pn1-\pn5-2\!\!\pn3) \: v_e(p_5)] \; 
[\bar{u}_{\chi{_i}}(p_4) \, G^{Aik}_{c 2} \, u_e(p_2)]}
{(p_1-p_5)^2 \, \{ (p_1-p_3-p_5)^2 - m_{\tilde{e}_k}^2 \}} \non \\ \non \\
M_{d 2} &=& -\biggl(\frac{e^3}{\stw} \biggr)\,
\frac{[\bar{v}_e(p_1) \, \gamma^\alpha \, v_e(p_5)] \; 
[\bar{u}_{\chi{_i}}(p_4) \, \gamma_\alpha \, (\pn1-\pn4-\pn5-\mgj)\; G^{Ajk}_{d 2} \, u_e(p_2)]}
{(p_1-p_5)^2 \, \{ (p_1-p_4-p_5)^2 - \mgj^2 \}} \non \\ 
\eeq
The relative signs among different diagrams contributing to the same
final state arising out of the anticommuting nature of the fermionic fields
(Wick's theorem) are summarized in the table below.  \s

\vskip 10pt
\begin{center}
\renewcommand{\arraystretch}{1.5}
\begin{tabular}{|c||c|c|c|c|c|c|c|c|c|c|c|c|}
\hline
Diagrams &  $a_1$ & $a_2$ & $a_3$ & $a_4$ & $a_5$ & $b_1$ & $b_2$ & $b_3$ & $b_4$ & $b_5$ &
$c_1$ & $c_2$ \\
\hline
\hline
Relative Signs & + & + & + & $-$ & + & $-$ & $-$ & $-$ & + & + & $-$ & $-$ \\
\hline
\end{tabular}
\end{center}

\nn We try to keep the conventions of our previous paper \cite{first} in
defining the variables and constants more or less intact. 
$P_{L,R}=\frac{1}{2}(1\mp \gamma_5)$ are the left-- and right--handed chirality
projectors, $e^2 = 4\pi \alpha$ with $\alpha$ being the fine structure
constant, $\stw=\sin\theta_W$ and $\ctw=\cos\theta_W$ are the sine and the
cosine of the Weinberg angle, while $T_{3,\tilde{e}_k(e)}$ and
$Q_{\tilde{e}_k(e)}$, are the third component of the weak isospin and the
charge of the $k$-th chiral selectron (electron).  The couplings of electron
and selectrons with gauge bosons are parametrized by
\beq
c_L^e = T_{3,e} - Q_e \stwn2 \qquad , \quad  c_R^e = - Q_e \stwn2 
\qquad , \quad  
a_{Z \tilde{e}_k \tilde{e}_k} = T_{3,\tilde{e}_k} - Q_{\tilde{e}_k} \stwn2
\eeq 
where $k=L,R$ is the chirality of the selectron, while the 
neutralino-neutralino-$Z$ boson couplings are parametrized by
\beq
O^L_{ij} = - O^R_{ij} =
-\frac{1}{2} N_{i3} N_{j3} + \frac{1}{2} N_{i4} N_{j4}
\eeq
following the notations of Gunion and Haber \cite{Haber}. As in
Ref.~\cite{first}, for the fermion-sfermion-gaugino couplings we follow
Figs.~22,23 and 24 of Re.~\cite{Haber}. In the amplitudes presented in
eqs.~(A1,A2), the couplings $G$ absorb the sign on imaginary $i$'s as shown
against the vertices in the figures mentioned above. The subscripts of $G$
indicate the diagram while superscripts ($i$ or $j$) indicate whether the
coupling is arising at a vertex with an outgoing ($i$) or a propagator ($j$)
gaugino. A superscript $k(h)=L,R$ indicates the chirality of the sfermion in
the final state (or in the scalar propagator) as appropriate for the vertex in
context. $G$-couplings with superscripts $A$ and $B$ are related by hermitian
conjugation at the Lagrangian level, while involving the same set of fields in
respective cases.  In all the couplings defined above we have taken the
neutralino-mixing matrix $N$ in the $B-W^3$ basis, the neutralino-mixing matrix
$N'$ in the $\tilde{\gamma}-\tilde Z$ basis, the two chargino mixing matrices
$U$ and $V$, to be real as is appropriate in an analysis that conserves CP. \s 

Following are the couplings in details:
\beq
G^{Ai(j)(L,R,\tilde{\nu})}_{(Diag.)} &=&C^{A}_{L(R)} \, (A^{i(j)}_{1(L,R,
\tilde{\nu})} P_L + A^{i(j)}_{2(L,R,\tilde{\nu})} P_R) \non \\
G_{(Diag.)}^{Bi(j)L(R)} &=& C^B_{L(R)} \, (B^{i(j)}_{1L(R)} P_R + 
B^{i(j)}_{2L(R)} P_L) 
\eeq
The generic values and structures of the terms on the right hand side of the 
above equations are defined below in reference to final states with neutralino 
and with chargino. For final states with neutralinos, one has:
\beq
 C^{A}_{L(R)} &=& C^{B}_{L(R)} = -1 \non \\ \non
A_{i(j)1L} &=& B_{i(j)1L} =
-\sqrt{2} \, \biggl[\stw N^\prime_{i(j)1} + \biggl(\frac{1}{2}-\stwn2\biggr)
N^\prime_{i(j)2}\biggr] \non \\
A_{i(j)2R} &=& B_{i(j)2R} =
\sqrt{2} (\stw N^\prime_{i(j)1} - \stwn2 N^\prime_{i(j)2}) \non \\
A_{i(j)2L} &=&  A_{i(j)1R} = B_{i(j)2L} =  B_{i(j)1R} = \frac{m_e}{\sqrt{2} m_W}
\biggl( \frac{N_{i(j)3}}{\cos\beta} \biggr)
\eeq 
[Note that we have  neglected the very small mixing in the selectron sector
although retained the small electron-mass ($m_e$) in the above couplings; the
reason behind this is discussed in Appendix B.] For final states with 
charginos, as in case for the Feynman diagrams in Fig. 3d, one has:
\beq
C^{A}_L &=& -C^{B}_L = +1  \qquad C^{A}_R =  C^{B}_R = 0 \non \\
A_{i(j)1} &=& = V_{i(j)1} \hskip -3pt \qquad \quad A_{i(j)2} =  -\frac{m_e}{\sqrt{2} m_W} 
\biggl(\frac{U_{i(j)2}}{\cos \beta}\biggr) \non \\
B_{jl} &=& A_{jl} \qquad {\rm with} \quad l=1,2 
\eeq
In the present analysis  we do not use the sparticle widths, and hence, have to
impose kinematic constraints to avoid situations closely approaching the
on--shell limits for the (sfermion and/or gaugino) propagators that would
have led to resonances. In Ref.~\cite{first} we have justified this
consideration and illustrated with concrete examples the importance of including
sparticle widths when approaching such thresholds.

\subsection*{Appendix B: the differential cross section}
\setcounter{equation}{0}
\renewcommand{\theequation}{B\arabic{equation}}

In this Appendix, we present the analytical expressions of the differential
cross section for the following two associated production processes: 
\beq
e^+(p_1) \; e^-(p_2) &\to& \tilde{e}_k^- (p_3) \; \chi_i^0(p_4) \; e^+(p_5)  \\
e^+(p_1) \; e^-(p_2) &\to& \tilde{\nu} (p_3) \; \chi_i^-(p_4) \; e^+(p_5)
\eeq
where $k=L,R$ is the handedness of the produced selectron. We have considered
the (gauge invariant) sets of only two Feynman diagrams, Fig. 3c and 3d, i.e.
for the processes which contain the photon pole in the $t$--channel. 
This is a reasonably good approximation, as demonstrated in Table 1, for the 
``full" cross section. \s

Due to the presence of forward poles in these diagrams, we have to keep the
electron mass in our analysis which ultimately regulates this singular behavior
of the scattering matrix element over the appropriate phase space.  However, it
is sufficient to retain the terms proportional to $m_e^2/t_1^2$, where  $m_e$
is the mass of the electron and $t_1$ is the square of the 4-momentum transfer
between the incoming and outgoing positron. Elsewhere it is safe to take $m_e
\to 0$. Hence, in the following expressions for the diagonal and interference
terms we have made two parts explicit; one proportional to $m_e^2/t_1^2$ and
the other, not. We ignored, however, the negligible mixing in the selectron
sector. \s

The complete spin-averaged matrix element squared is given by
\begin{equation}
|{\cal M}|^2 = 4 \frac{(4\pi \alpha)^3}{s_W^2} \; \sum_{i,j=1,2} T_{ij}
\end{equation}
where $\alpha$ is the fine structure constant [to be evaluated at the scale of
the momentum transfer at the vertex]. The $T_{ij}$'s are the squared
amplitudes and the interferences for the two diagrams in Figs. 3c (and 3d). The
effects of permutations of suffixes for the interferences are already included
in the original terms and hence such permutations are to be left out. \s

The diagonal terms are given by:
\beq
T_{c_1(d_1)} = 
\frac{1}{s_2^2} \;
\Biggl[  
\frac{A_{i1k}^2}{t_1^2} \,
\Biggl( 
\cala1 + \cala2 + \cala3
\Biggr) - 
  \frac{m_e^2}{t_1^2} \: 
\Biggl\{
A_{i1k}^2 \cala4
 - \frac{A_{i2k}^2}{m_W^2} \, (\cala1 + \cala2)
\Biggr\} \Biggr] \non
\eeq
\beq
T_{c_2} &=&
\frac{-2}{(m_{\chi_i}^2 - s_2 + t_1 - t_2)^2}
\Biggl[ \frac{A_{i1k}^2}{t_1^2} 
\calb1 \biggl(\calb2 \calb3 + \msf^2 t_1 \biggr) \non \\
&&-  
\frac{m_e^2}{t_1^2}
\Biggl\{
4 A_{i1k} \biggl( \frac{A_{i2k}}{m_W} \biggr) \calb3 + 
\biggl\{
\biggl( \frac{A_{i2k}}{m_W}\biggr)^2 \calb3 - A_{i1k}^2  
\biggr\}
\calb4 
\Biggr\}
\Biggr] \non  
\eeq
\beq
T_{d_2} &=& 
\frac{1}{(t_2 - \mgj^2)^2} \,  
\Biggl[
\frac{A_{j1k}^2}{t_1^2} \,
\Biggl\{
\calc1 + (\calc2 + \calc3 + \calc4 + \calc5 + \calc6) \, t_1
\Biggr\} \non \\ && -
\frac{m_e^2}{t_1^2} \,
\Biggl\{
 8 A_{j1k} \, \biggr(\frac{A_{j2k}}{m_W} \biggr) \, \calc7 - 
\biggr(\frac{A_{j2k}}{m_W} \biggr)^2 \, \calc5 - 
A_{j1k}^2 (\calc2 + 2 \calc3 - 2 \calc4 - 3 \calc5 + \calc8)
\Biggr\}
\Biggr] \ \ \ 
\eeq
while the interference terms are given by:
\beq
&& T_{c_1c_2} = 
\frac{1}{s_2 (m_{\chi_i}^2 - s_2 + t_1 -t_2)}
\Biggl[  
\frac{A_{i1k}^2}{t_1^2} 
\Biggl\{  \cald1 + \cald2 
+ m_{\chi_i}^2 (\cald3 + \cald4)
+ m_{\tilde{f}}^2 (\cald5 + \cald6) 
\Biggr\} 
\non  \\  
&&-   
\frac{m_e^2}{t_1^2} 
\Biggl\{ 
A_{i1k} 
\Biggl( 
2 \frac{A_{i2k}}{m_W} m_{\chi_i} \cald7- A_{i1k} \cald8 
\Biggr) 
+ \frac{A_{i2k}}{m_W}
\Biggl(
2 m_{\chi_i} A_{i1k} \cald7 + \frac{A_{i2k}}{m_W} (\cald1+\cald3+\cald5)
\Biggr)
\Biggr\} 
\Biggr] \non 
\eeq
\beq
&& T_{d_1d_2} = 
\frac{-1}{s_2 \, (m_{\chi_j}^2 - t_2)} \,
\Biggl[  
\frac{A_{i1k} A_{j1k}}{t_1^2} \;
\Biggl\{ 
\calf1 + \calf2 
+ m_{\chi_i}^2 (\calf3 + \calf4)
+ m_{\tilde{f}}^2 (\calf5 + \calf6) 
\Biggr\} \non  \\  
&&  -   
\frac{m_e^2}{t_1^2} 
\Biggl\{ 
A_{i1k} 
\Biggl( 
2 \frac{A_{j2k}}{m_W} m_{\chi_i} \calf7- A_{j1k} \calf8 
\Biggr) 
+ \frac{A_{i2k}}{m_W}
\Biggl(
2 m_{\chi_i} A_{j1k} \calf7 + \frac{A_{j2k}}{m_W} (\calf1+\calf3+\calf5)
\Biggr)
\Biggr\} 
\Biggr] \non \\
&&  
\eeq
\nn The factors $\cal{A}_i$, $\cal{B}_i$, $\cal{C}_i$, $\cal{D}_i$, 
are given below: 
\beq
\cala1 &=& s \, \left\{ (\mgi^2-\msf^2) \, (2 s - s_2) + \msf^2 \: s_2 \right \} \non \\ 
\cala2 &=& s_2 \, \biggl[ s \left\{ 2 (s - s_1 - s_2) + t_2 \right \} 
        + s_2 \, (s_1 - t_2) \biggr] \non \\ 
\cala3 &=& t_1 \left\{ (\mgi^2-\msf^2) \, (2 s - s_2 + t_1)
       + s_2 (s - s_1 + t_2) \right\} \non \\ 
\cala4 &=& 6 \,s \, (\mgi^2 - \msf^2) + s_2 \left\{ 4 \, (\msf^2 + s) - 3 \,(s_1 + s_2) - t_2 
           \right\} \non\\ \non\\
\calb1 &=& \msf^2 - s_2 + t_1 -t_2 \non \\ 
\calb2 &=& \mgi^2 + s - s_1 - s_2 \non \\ 
\calb3 &=& s - s_1 + t_2 \non \\ 
\calb4 &=& - \msf^2 + s_2 + t_2 \non \\ \non \\
\calc1 &=& -2 \, \msf^2  s_1^2  + s_1 \biggl\{ 2 \, (\msf^2  - s + s_1) + s_2 \biggr\} \, t_2 
  + (s - 2 s_1 - s_2) \, t_2^2  \non \\
 &&  + \mgj^2 \, (2 s s_1 - s_1 s_2 - s t_2 + s_2 t_2) 
  + \mgi^2  \, \biggl\{ - \mgj^2  s + 2 \msf^2  (s_1 - t_2) + t_2 (s - 2 s_1 + 2 t_2) \biggr\}
\non \\ 
\calc2 &=& - \mgj^2 \, (s_2 + t_2) + s_2 t_2 \non \\ 
\calc3 &=& - 2 \, \mgi \, \mgj (\msf^2 - t_2)  \non \\ 
\calc4 &=& \mgj^2 \, s - \msf^2 \, (2 s_1 - t_2) - s t_2 \non \\ 
\calc5 &=& s_1 \, (\mgj^2 + t_2) \non \\ 
\calc6 &=& \mgi^2 \, \msf^2 - (\mgi^2 - \mgj^2 + \msf^2) \, t_1 \non \\ 
\calc7 &=&  \mgj \, (\mgi^2 - s_1) \, (s_1 - t_2) \non \\ 
\calc8 &=&  2 \, \biggl\{ \mgj^2 \,(\mgi^2  + \msf^2 ) - \mgi^2  s_1 + s_1^2 \biggr\} + t_2^2 
 \non \\ \non \\
\cald1 &=& \mgi^4  s + 2 \, s^2 \, (2 s_2 + t_2) + s_2 \, (s_1 - t_2) \, (2 s_1 + 3 s_2 + t_2) 
\non \\
     && + \, s \, \left\{-4 s_2^2  + s_2 t_2 + t_2^2  - 2 s_1 \, (3 s_2 + t_2) \right\} \non \\
\cald2 &=& -t_1 \left\{2 \msf^4  + (s - s_1) \, (2 s - 2 s_1 - 3 s_2) + (s - s_1 - 2 s_2) \, 
           t_2 \right\} \non \\
\cald3 &=& -2 \msf^2 \, s + (2 s + s_2) \: (s - s_1 + t_2) \non \\
\cald4 &=& t_1 \: (\msf^2  - s + s_1 - 2 \, t_2) \non \\
\cald5 &=& -4 s^2  + 4 s (s_1 + s_2) - 2 s t_2 - 2 s_2 \, (s_1 - t_2) \non \\
\cald6 &=& t_1 \: (-2 s + 2 s_1 + 3 s_2 - 2 t_1 + t_2) \non \\
\cald7 &=& \mgi^2  s + 2 s^2  + s_2 (s_1 - t_2) - s \left\{2 \, (s_1 + s_2) - t_2 \right\} 
\non \\
\cald8 &=& 2 \mgi^4  + 2 s s_1 - 2 s_1^2  + 6 s \, s_2 - 6 s_1 s_2 - 5 \, s_2^2  + 3 \, s t_2 
\non \\
   && - \, s_1 t_2 - s_2 t_2 + t_2^2  + \mgi^2  (-4 \msf^2  + 3 s - s_1 + s_2 + t_2) \non \\
   && - \, 2 \msf^2  \left\{ 4 s - 3 (s_1 + s_2) + t_2 \right\} \non \\ \non \\
\calf1 &=& -\mgi^4  s + 4 \mgi \mgj s (s + s_2) - s_1 s_2 (-2 s + 2 s_1 + s_2) \non \\
       &&  + (2 s + s_2) (-s + s_1 + s_2) t_2  + (-s + s_2) t_2^2 \non \\
\calf2 &=& t_1 \{2 \msf^4  + 2 \mgi \mgj (2 s - s_2 + t_1) 
        + (s - s_1) (2 s - 2 s_1 - s_2 + t_2)\} \non \\
\calf3 &=& 2 \msf^2 \, s - 2 s^2  + s_2 \, (s_1 - t_2) + 2 s \, (s_1 + s_2 - t_2) \non \\
\calf4 &=& t_1 \, (-\msf^2 - s_1 + s + 2 \, t_2) \non \\
\calf5 &=& -2 \, \{ (s_1 - t_2) \, (s - s_2) \, + s s_1 \} \non \\
\calf6 &=& -t_1 \, \{2 \, (s + s_1) \, + (s_2 + t_2) \, \} \non \\
\calf7 &=& - \, \{ s (\mgi^2 - s_1) - (s-s_2) \, (s_1-t_2) \} \non \\
\calf8 &=& -2 \mgi^4  - 2 s s_1 + 2 s_1^2  + 4 \mgi \mgj (3 s - s_2) + 2 s s_2 - s_2^2 \non \\
      && + \mgi^2 (4 \msf^2  - 3 s + s_1 + 3 s_2 - t_2) - 3 s t_2 + s_1 t_2 - s_2 t_2 - t_2^2 \non \\
       &&  + 2 \msf^2  (-2 s - 3 s_1 + s_2 + t_2) \hspace*{-2cm}
\eeq


\nn where the 5 Lorentz--invariant Byckling-Kajantie (Mandelstam-analogue) 
\cite{kajantie} variables for a $(2 \to 3)$ process are given (in terms of 
4--momenta) by
\beq
s = (p_1+p_2)^2  \quad 
s_1 = (p_4+p_5)^2   \quad
s_2 = (p_3+p_4)^2   \quad
t_1 = (p_1-p_5)^2   \quad
t_2 = (p_2-p_3)^2   
\eeq
$A_{ink}$ are the couplings of the $i$-th gaugino with a sfermion of chirality
$k$ and a fermion. $n=1,2$, where 1 gives the coupling with SU(2) and U(1)
gaugino content of the eigenstate and 2 gives that with the higgsino part and
proportional to the electron mass.  $A_{i(j)nk}=A_{i(j)n}$ for $\tilde \nu$
final states as in Fig. 3d. 
Also we have used scaled $A_{i(j)2k}$ and $A_{i(j)2}$
in Appendix B which are equal to $A_{i(j)2k}$ in eq.(A6) and $A_{i(j)2}$ 
in eq.(A7), respectively,  but divided by $m_e$. This is done to make the
$m_e$ dependences of the squared-amplitudes explicit which is important
for our purpose (as discussed in the beginning of this appendix), rather
than keeping it concealed under the definitions of the couplings.
  Note that the mass-dimensions of the
variables ${\cal A}_i$, ${\cal B}_i$, ${\cal C}_i$, ${\cal D}_i$ and ${\cal
F}_i$ are left arbitrary to render the expressions for the squared terms
$T_{ij}$'s to have a somewhat systematic look. \s

The differential cross section is obtained by dividing by the flux and
multiplying by the phase space,
\beq
{\rm d} \sigma = \frac{1}{2s} \times \frac{1}{(2 \pi)^5} \frac{ {\rm d}^3p_3}{
2E_3} \frac{ {\rm d}^3p_4}{ 2E_4} \frac{ {\rm d}^3p_5}{ 2E_5} \delta^4 (p_1+p_2
-p_3-p_4-p_5) \times |M|^2
\eeq
The integral over the phase space, to obtain the total production cross
section, is then performed numerically. \s

Finally, we note that the production of the charged conjugate states has also
to be taken into account. Due to CP--invariance, these cross sections are the
same as for the corresponding previous ones, which have thus to be multiplied
just by a factor of two.


\begin{thebibliography}{99} 

\bibitem{first} A. Datta and A. Djouadi, hep-ph/0111466. 
%
\bibitem{egamma1} F.M. Renard,  Z.~Phys.~C14 (1982) 209;
J.~A.~Grifols and R.~Pascual, Phys.~Lett.~B142 (1984) 455 and B135 (1984) 319;
M. Drees and K. Grassie, Z.~Phys.~C28 (1985) 451.
%
\bibitem{egamma2}
F.~Cuypers, G.~van Oldenborgh and R.~Ruckl, Nucl.~Phys.~B409 (1993) 144;
D.~Choudhury and F.~Cuypers, Nucl.~Phys.~B451 (1995) 16;
S.~Hesselbach and H.~Fraas, Phys.~Rev.~D55 (1997) 1343;
V.~Barger, T.~Han and J.~Kelly, Phys.~Lett.~B419 (1998) 233;
C.~Blochinger and H.~Fraas, Acta Phys.~Polon.~B30 (1999) 3417.
%
\bibitem{WW} C.F.~Weizs\"acker, Z.~Phys.~C88 (1934) 612; 
E.J.~Williams, Phys.~Rev.~D45 (1934) 729. 
%
\bibitem{IWW} S. Frixione, M.L. Mangano, P.~Nason and G.~Ridolfi, Phys.~Lett. 
B319 (1993) 339. 
%
\bibitem{laser} I.\, Ginzburg,  G.\, Kotkin, V.\, Serbo and V.\, Telnov, 
Nucl. Instrum. Meth. 205 (1983) 47 and 219 (1984) 5; 
V.\, Telnov, Nucl. Instrum. Meth. A294 (1990) 72 and A335 (1995) 3;
J. K\"uhn, E. Mirkes and J. Steegborn, Z. Phys. C57 (1993) 615.
%
\bibitem{Haber} J.\, Gunion and H.\, Haber, Nucl. Phys. B272 (1986) 1, (E) 
hep-ph/9301205.

\bibitem{PO} For the couplings see also A. Djouadi, J. Kalinowski, P. 
Ohmann and P.M. Zerwas, Z.~Phys.~C74 (1997) 93 and 
P.M. Zerwas (ed.) et al., hep-ph/9605437. 
%
\bibitem{newgamma} ECFA/DESY Photon Collider Working Group, DESY-2001-011, 
hep-ex/0108012.
%
\bibitem{NLC} E. Accomando, Phys. Rept. 299 (1998) 1;  American Linear Collider
Working Group (T. Abe et al.),  Report SLAC-R-570 and hep-ex/0106057; J. Bagger
et al., hep-ex/0007022; H. Murayama and M. Peskin, Ann. Rev. Nucl. Part. Sci. 
46 (1996) 533; TESLA TDR, Part III: ``Physcis at $\ee$ Linear Collider"
D. Heuer, D. Miller, F. Richard and P.M. Zerwas (eds.) et al., Report 
DESY--01--011C, hep-ph/0106315.  
%
\bibitem{LEP2} For a summary on the experimental limits of the masses of
the Higgs and SUSY particles, see F. Gianotti, talk given at the 
EPS--HEP--2001, 12--18 July, Budapest. 
%
%
\bibitem{Comphep} 
 A. Pukhov, E. Boos, M. Dubinin, V. Edneral, V. Ilyin, D.
Kovalenko, A. Kryukov, V. Savrin, S. Shichanin, and A. Semenov,  ``CompHEP - 
a package for evaluation of Feynman diagrams and integration over 
multi-particle phase space. User's manual for version 33", Preprint INP MSU 
98-41/542, hep-ph/9908288. 
%
\bibitem{alpha} E. Boos and T. Ohl, Phys.~Rev.~Lett.~83 (1999) 480.
%
\bibitem{poles} See for instance: G. Passarino,  hep-ph/9810416; 
E.E. Boos and M.N. Dubinin, hep-ph/9909214.
%
\bibitem{Peter} For some examples of such studies, see: 
A. Freitas, D.J. Miller and P.M. Zerwas, Eur. Phys. J. C21 (2001) 361;
J.L. Feng and M.E. Peskin, Phys. Rev. D64 (2001) 115002. 
%
\bibitem{kajantie} E. Byckling and K. Kajantie, Phys. Rev. D187 (1969) 2008;
F.M. Renard, in ``Electron Positron Collisions", ed. Editions Fronti\`ers, 
Gif-sur-Yvette, 1981.
\end{thebibliography}
\end{document}